\documentclass[10pt,journal,compsoc]{IEEEtran}
\pdfoutput=1

\ifCLASSOPTIONcompsoc
  \usepackage[nocompress]{cite}
\else
  \usepackage{cite}
\fi
\ifCLASSINFOpdf
\else
\fi
\usepackage{amsmath}
\usepackage{amsmath}
\usepackage{hyperref}

\usepackage{booktabs}
\usepackage{colortbl}
\usepackage{graphicx}
\usepackage{subcaption}
\usepackage{changepage}
\usepackage{wrapfig}
\usepackage{algorithm}
\usepackage{algpseudocode}
\usepackage[colorinlistoftodos,prependcaption,textsize=tiny,textwidth=0.55in]{todonotes}

\definecolor{WHITE}{gray}{1.0}
\definecolor{GRAY}{gray}{0.9}

\newcommand{\editBegin}{}
\newcommand{\editEnd}{}

\newcommand{\conjac}{\textsc{ConJac}}
\newcommand{\vanilla}{\textsc{Vanilla}}

\newcommand{\SNH}{{\small{SNH}}}
\newcommand{\aSTVK}{{\small{aSTVK}}}
\newcommand{\aFUNG}{{\small{aFUNG}}}

\newcommand{\bunny}{\textsc{Bunny}}
\newcommand{\twist}{\textsc{Twist}}
\newcommand{\hetero}{\textsc{Hetero}}
\newcommand{\aniso}{\textsc{Aniso}}
\newcommand{\muscle}{\textsc{Muscle}}
\newcommand{\barCut}{\textsc{BarCut}}
\newcommand{\dragon}{\textsc{Dragon}}
\newcommand{\arma}{\textsc{Armadillo}}

\newcommand{\eg}{\textit{e.g.,}~}
\newcommand{\ie}{\textit{i.e.,}~}

\newcommand{\Eneg}[2]{${#1}\textsc{e-}{#2}$}
\newcommand{\Epos}[2]{${#1}\textsc{e}{#2}$}

\newcommand{\KK}{{\bf K}}
\newcommand{\JJ}{{\bf J}}
\let\SS\undefined 
\newcommand{\SS}{{\bf S}}

\newcommand{\Kdd}{\KK_{dd}}
\newcommand{\Kdq}{\KK_{dq}}
\newcommand{\Kqd}{\KK_{qd}}
\newcommand{\Kqq}{\KK_{qq}}
\newcommand{\Jqd}{\JJ_{qd}}

\pdfpageattr {/Group << /S /Transparency /I true /CS /DeviceRGB>>}

\let\orgautoref\autoref

\def\secnospace~{\S{}}
\renewcommand{\autoref}
        {\def\equationautorefname{Eq.}%
         \def\figureautorefname{Fig.}%
         \def\subfigureautorefname{Fig.}%
         \def\algorithmautorefname{Alg.\@}%
         \def\Itemautorefname{Item}%
         \def\tableautorefname{Table}%
         \def\sectionautorefname{\secnospace}%
         \def\subsectionautorefname{\secnospace}%
         \def\subsubsectionautorefname{\secnospace}%
         \def\chapterautorefname{\secnospace}%
         \def\partautorefname{Part}%
         \orgautoref}

\def\BibTeX{{\rm B\kern-.05em{\sc i\kern-.025em b}\kern-.08em
    T\kern-.1667em\lower.7ex\hbox{E}\kern-.125emX}}

\let\vv\undefined 
\newcommand{\vv}{{\bf v}}

\newcommand{\II}{{\bf I}}

\newcommand{\FF}{{\bf F}}
\newcommand{\RR}{{\bf R}}
\newcommand{\UU}{{\bf U}}

\newcommand{\VV}{{\bf V}}
\newcommand{\DD}{{\bf D}}

\newcommand{\ff}{{\bf f}}

\newcommand{\xx}{{\bf x}}

\newcommand{\bb}{{\bf b}}

\newcommand{\nn}{{\bf n}}

\newcommand{\MM}{{\bf M}}

\newcommand{\teaserHeight}{2.25cm}

\newcommand{\figDragon}{
  \begin{figure*}[t]
    \centering
    \includegraphics[height=\teaserHeight,trim={0cm 0 0cm 0},clip]{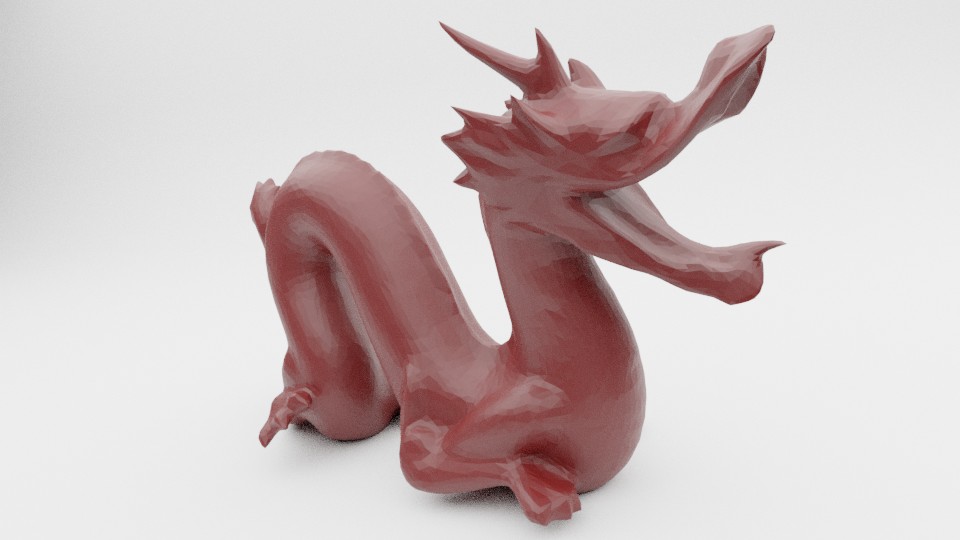}
    \includegraphics[height=\teaserHeight,trim={0cm 0 0cm 0},clip]{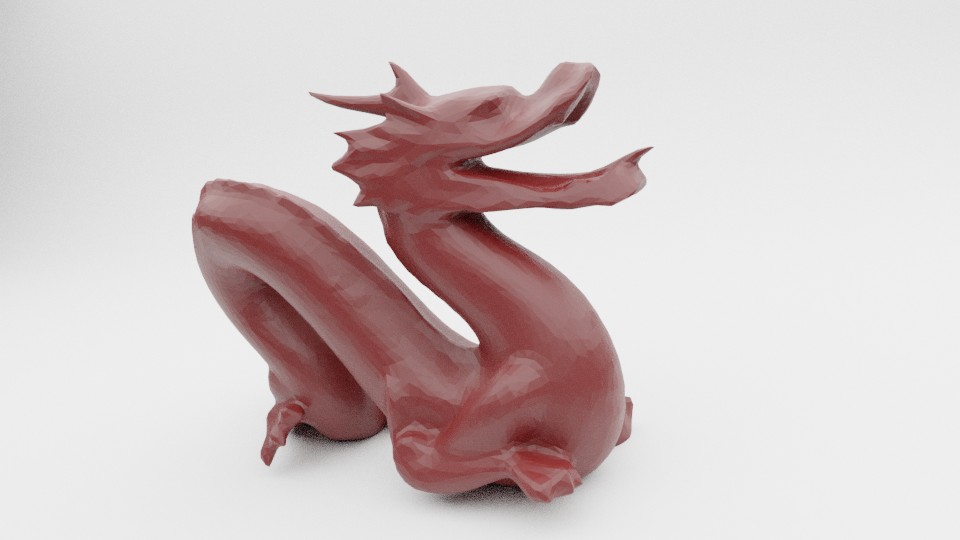}
    \includegraphics[height=\teaserHeight,trim={0cm 0 0cm 0},clip]{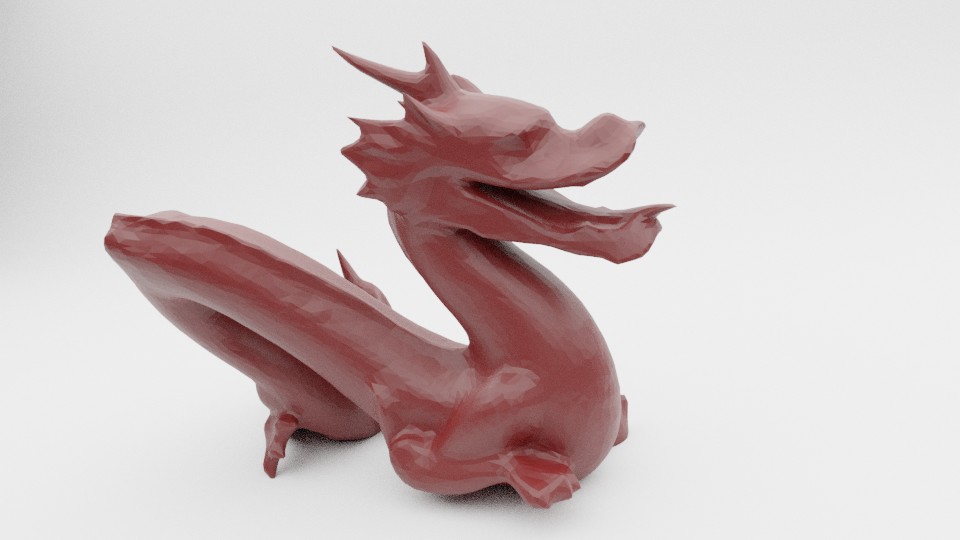}
    \includegraphics[height=\teaserHeight,trim={0cm 0 0cm 0},clip]{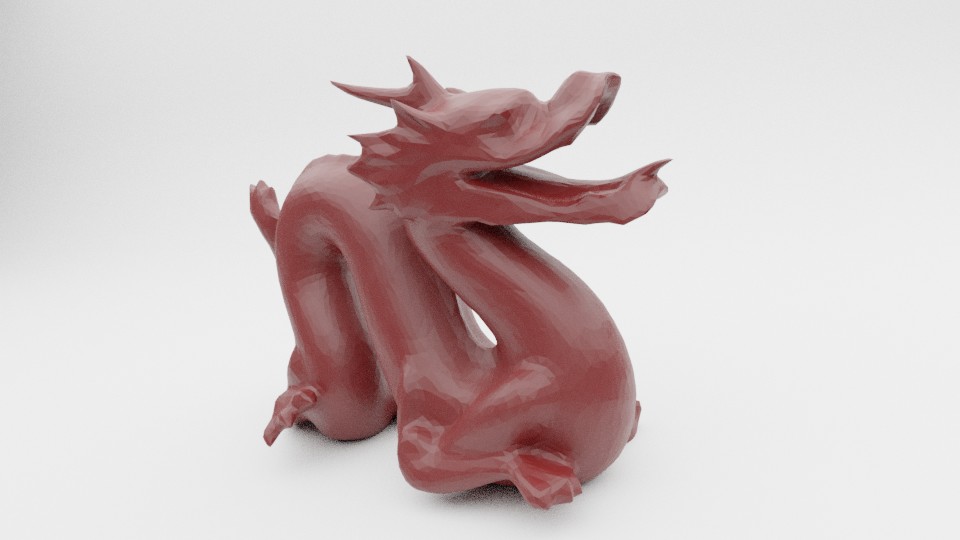}
    \caption{
      \dragon{}. Pulling portions of the mesh generates lively motion using only a small number of dynamic nodes. 
    }
	\label{fig:teaser}
  \end{figure*}
}

\newcommand{\figExtCost}{
  \begin{figure}[tb]
    \centering
    \includegraphics[width=\columnwidth]{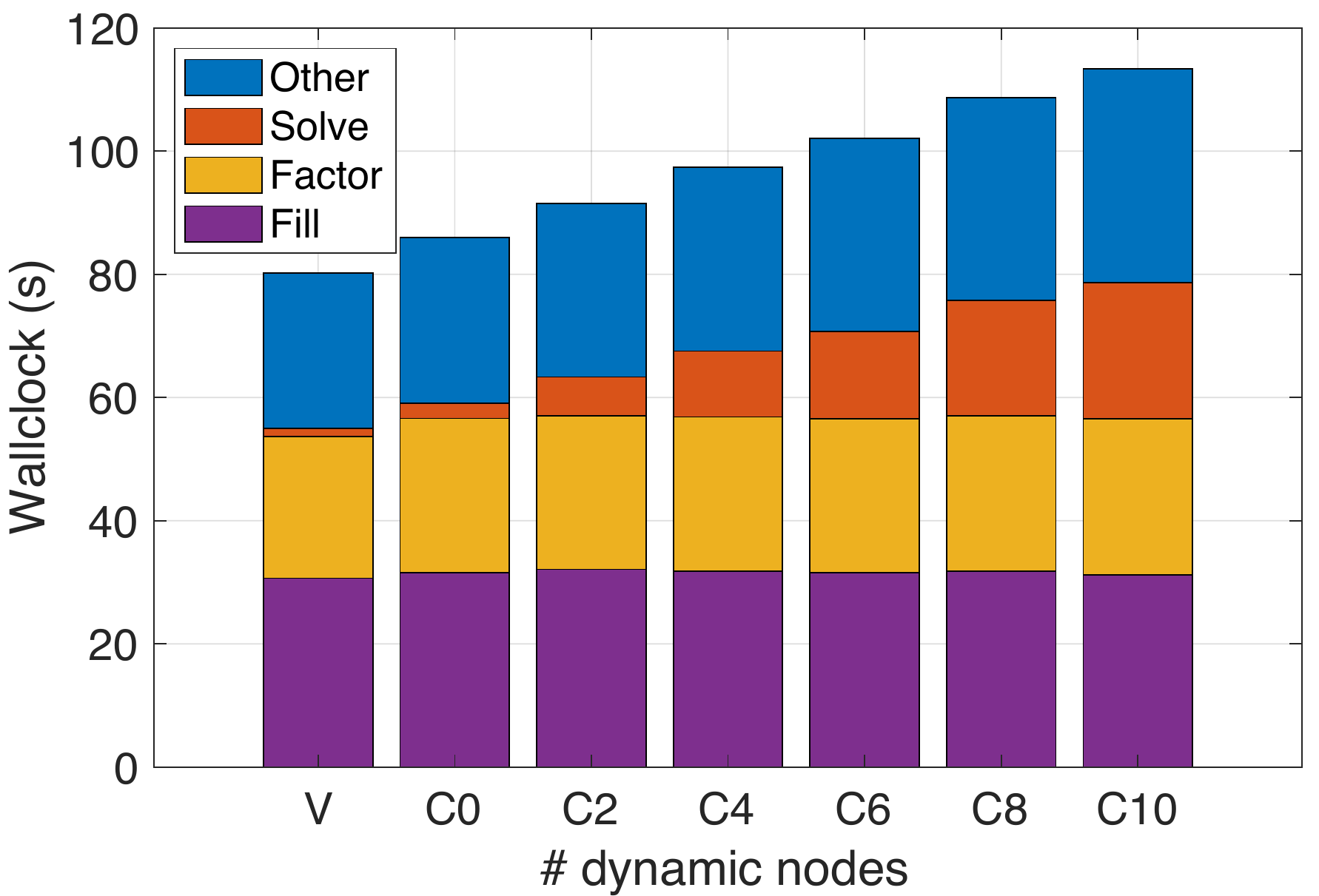}
    \caption{
      Wallclock times for \dragon{}. Starting from the left: \vanilla{}, \conjac{} with 0, 2, 4, 6, 8, and 10 dynamic nodes. Each bar is broken down into Fill ($\ff$ and $\KK$), Factor, Solve, and Other. As the number of dynamic nodes increases, the Solve cost goes up linearly.
    }
    \label{fig:ecost}
  \end{figure}
}

\newcommand{\figTwistInset}{
\begin{wrapfigure}{r}{0.9in}
  \vspace{-0.15in}
  \hspace{-0.15in}
  \includegraphics[width=0.99in]{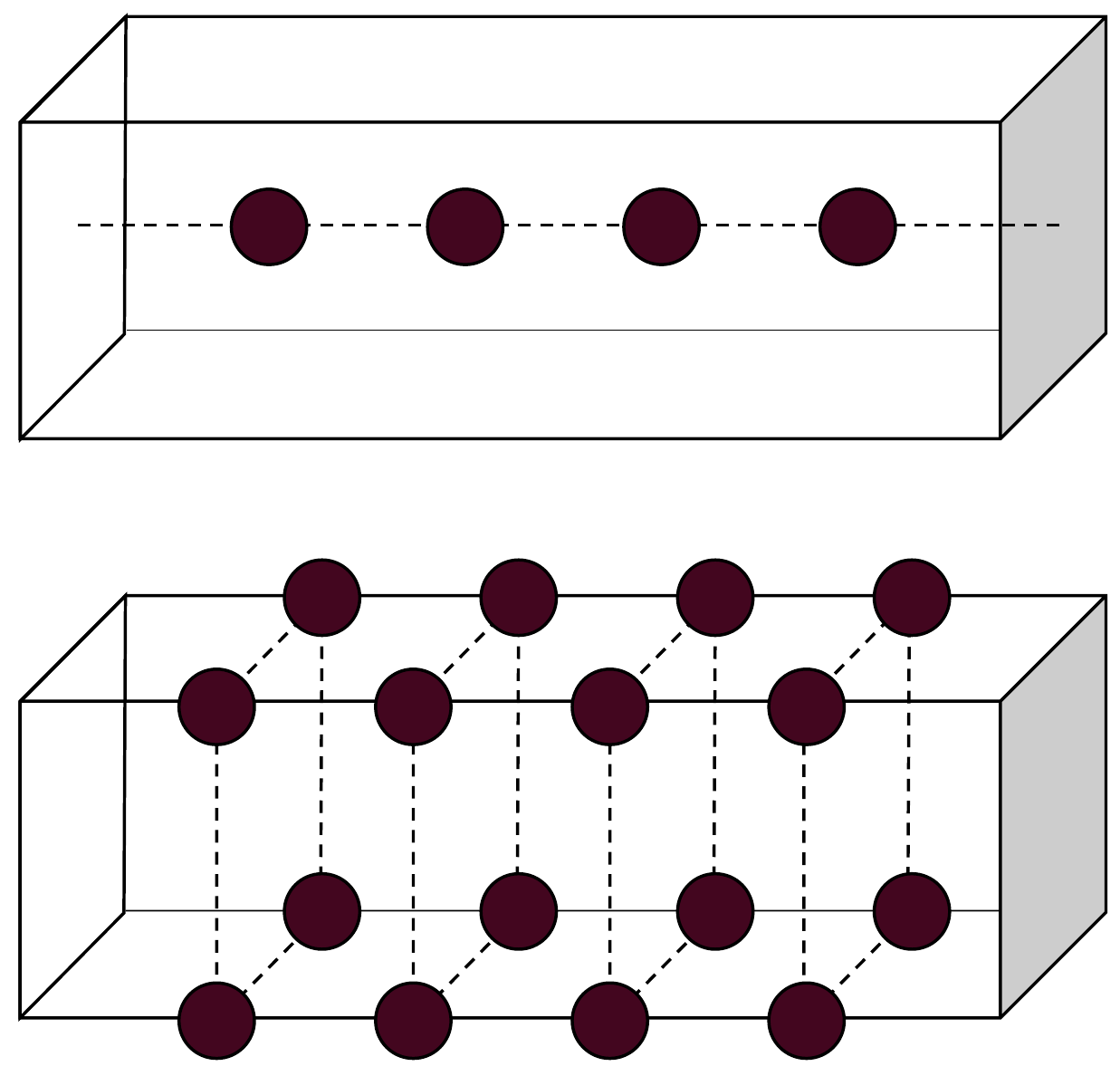}
  \vspace{-0.11in}
\end{wrapfigure}
}

\newcommand{\mysize}{1.7cm}
\newcommand{\figTwist}{
	\begin{figure}[b]
		\captionsetup[subfigure]{}
		\centering
		\subcaptionbox{\label{fig:exts1}}{
			\includegraphics[height=\mysize, trim=0 2.1cm 0 0.9cm, clip]{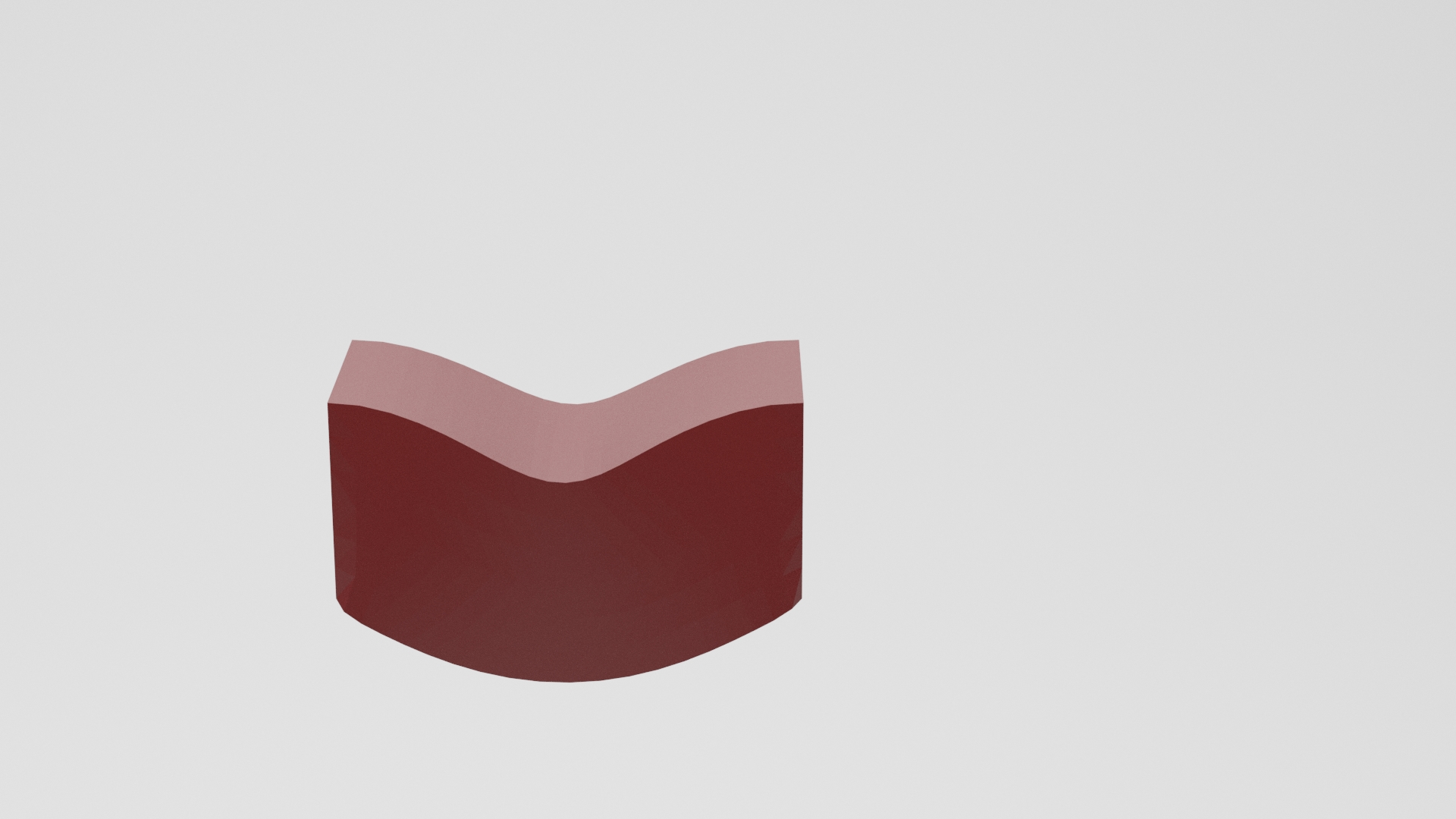}
		}
		\subcaptionbox{\label{fig:exts2}}{
			\includegraphics[height=\mysize, trim=0 2.1cm 0 0.9cm, clip]{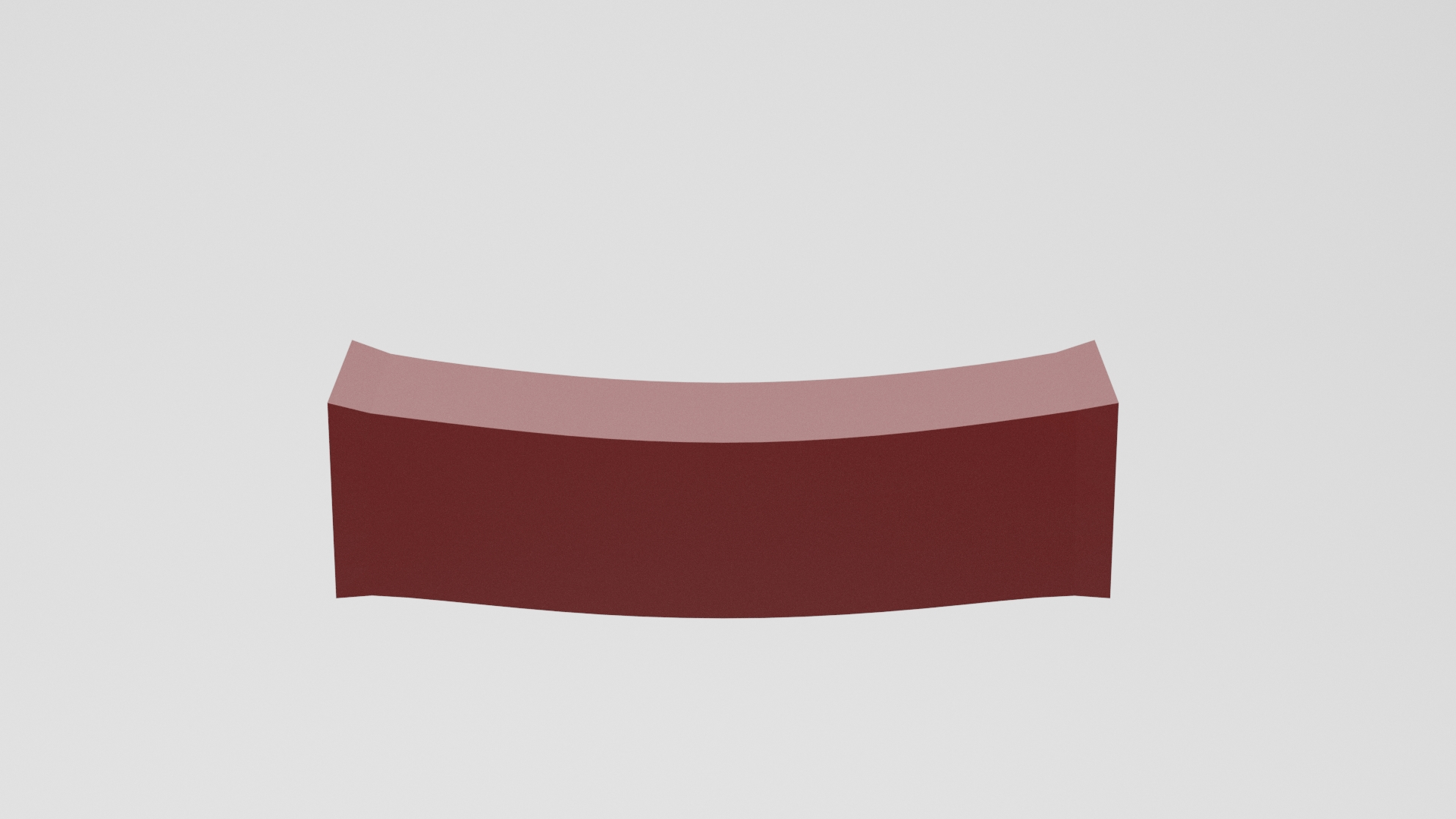}
		}
		\subcaptionbox{\label{fig:exts3}}{
			\includegraphics[height=\mysize, trim=0 2.1cm 0 0.9cm, clip]{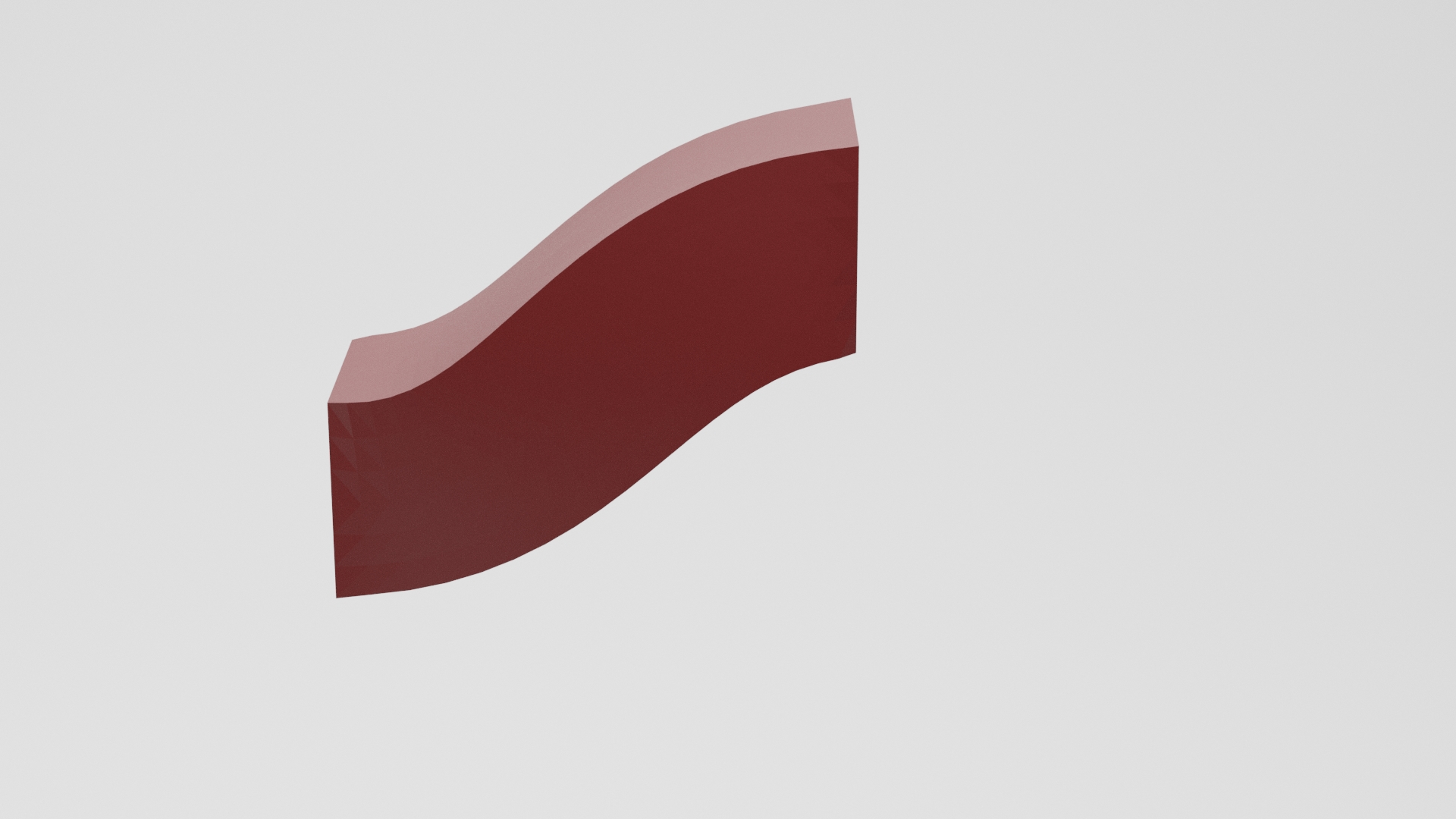}
		}
		\subcaptionbox{\label{fig:exts4}}{
			\includegraphics[height=\mysize, trim=0 2.1cm 0 0.9cm, clip]{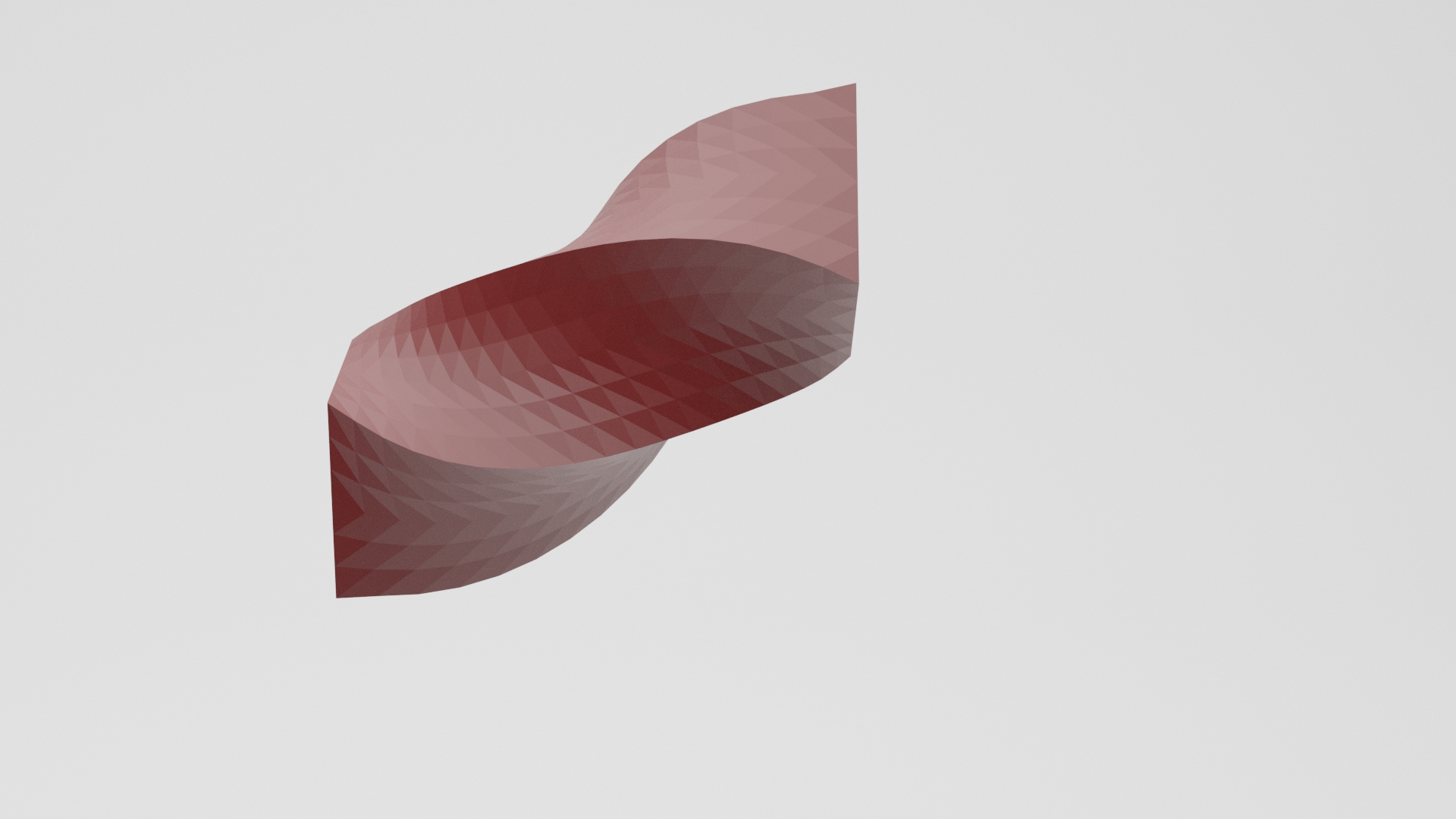}
		}
		\vspace{-0.15in}
		\caption{
			\twist{} scene: A bar is (a) compressed, (b) stretched, (c) bent, (d) and twisted through kinematic motion. 
			The bar in these figures only have a single dynamic node.
		}
		\label{fig:exts}
	\end{figure}
}

\newcommand{\figHetero}{
	\begin{figure}[b]
		\centering
		\subcaptionbox{\label{fig:hetero1}}{
			\includegraphics[height=\mysizeA]{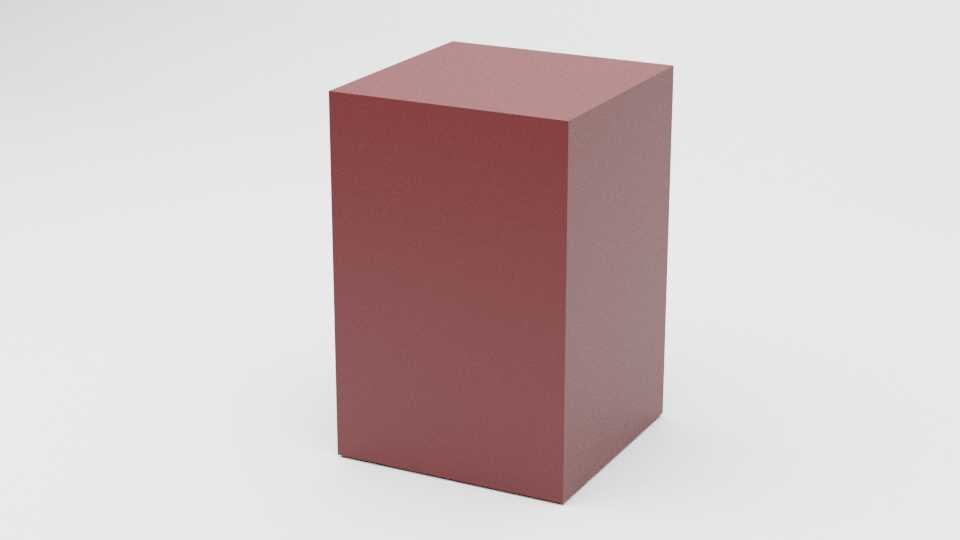}
		}
		\subcaptionbox{\label{fig:hetero2}}{
			\includegraphics[height=\mysizeA]{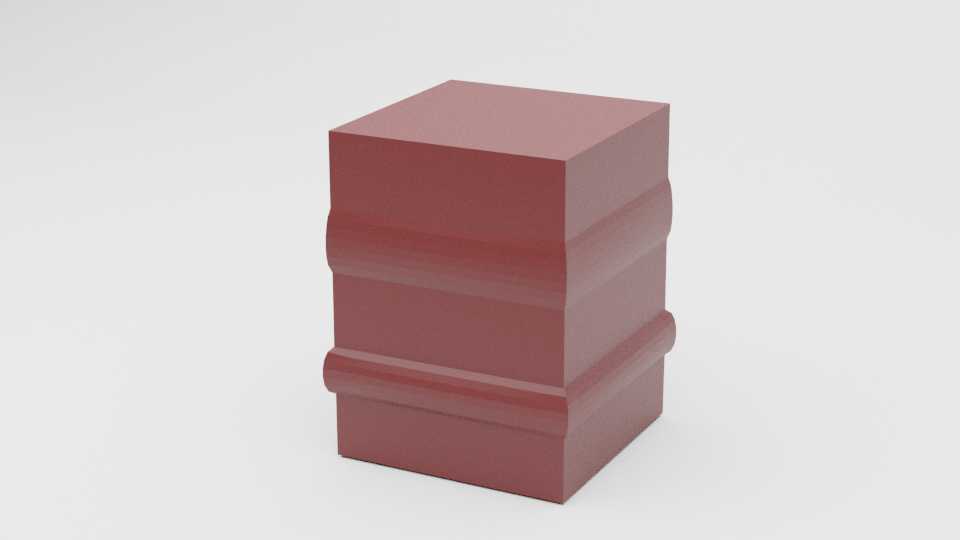}
		}
		\vspace{-0.15in}
		\caption{
			\hetero{} scene that divides a vertical bar mesh into layers of alternating material stiffness. 
		}
		\label{fig:hetero}
	\end{figure}
}

\newcommand{\mysizeB}{3.5cm}
\newcommand{\figAniso}{
	\begin{figure}[t]
		\captionsetup[subfigure]{}
		\centering
		\subcaptionbox{\label{fig:anisohor}}{
			\includegraphics[height=\mysizeB,trim={20cm 0 20cm 0},clip]{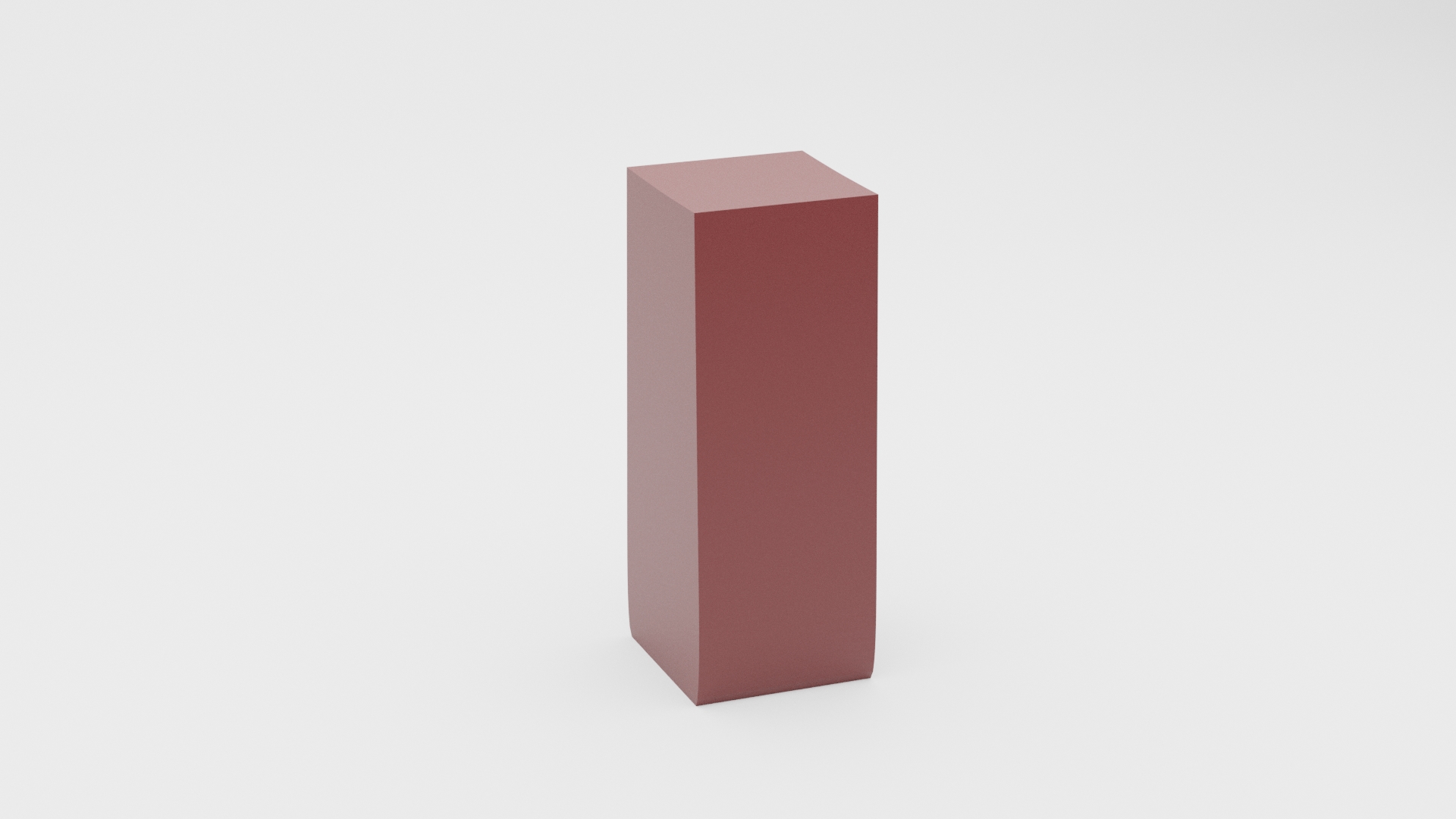}
		}
		\subcaptionbox{\label{fig:anisover}}{
			\includegraphics[height=\mysizeB,trim={20cm 0 20cm 0},clip]{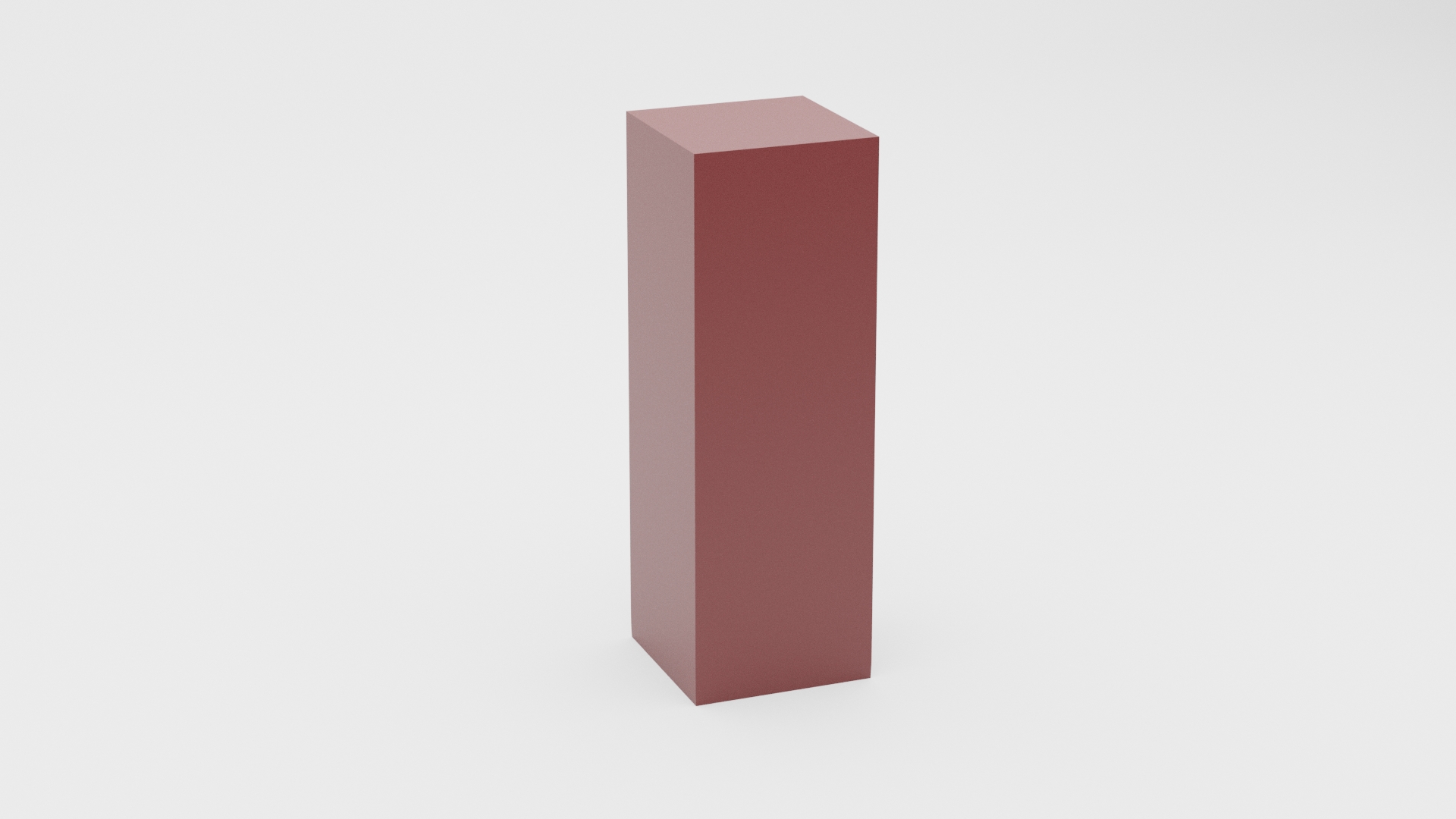}
		}
		\subcaptionbox{\label{fig:anisohel}}{
			\includegraphics[height=\mysizeB,trim={20cm 0 20cm 0},clip]{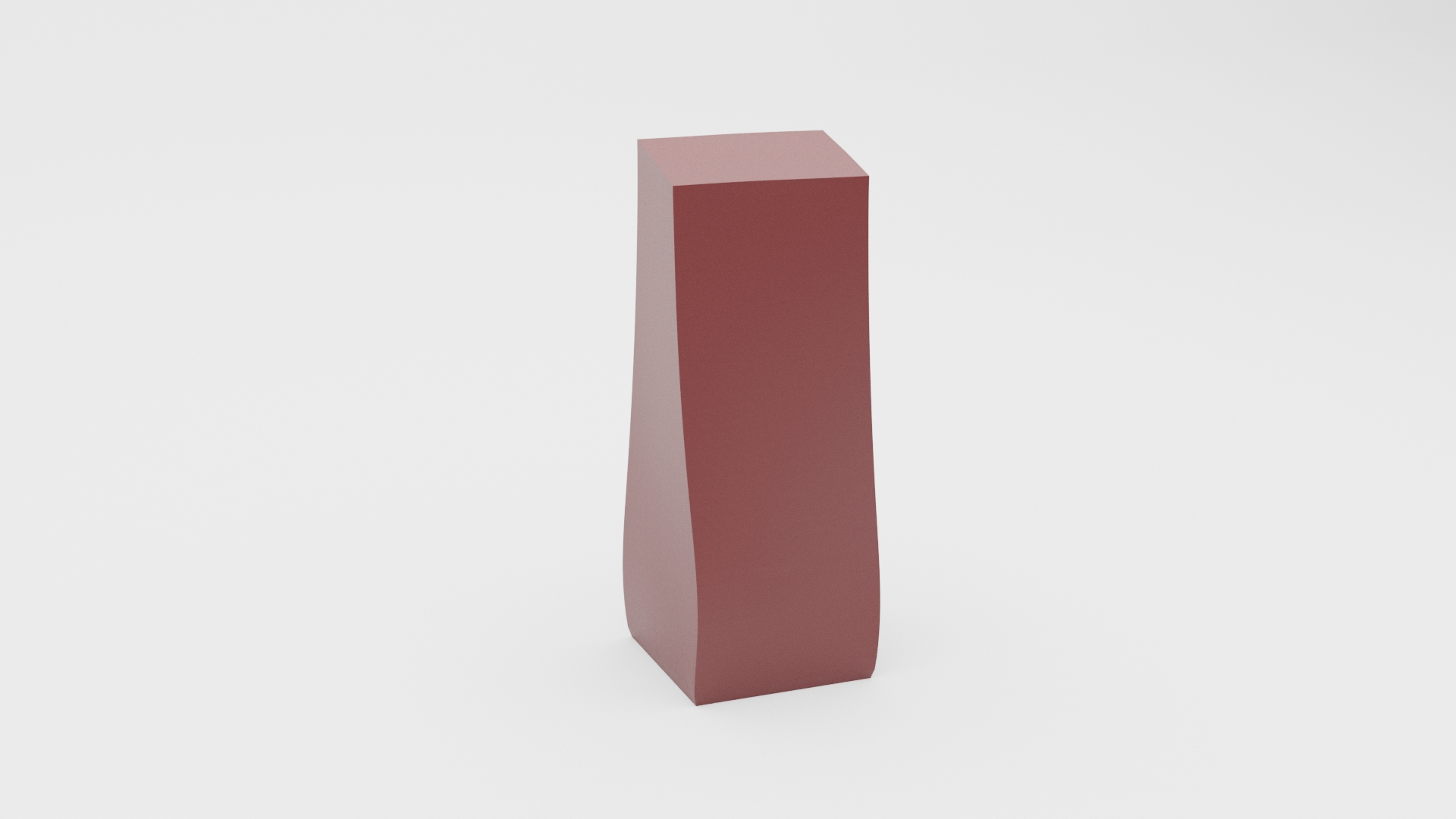}
		}
		\vspace{-0.15in}
		\caption{
			\aniso{} bar with (a) horizontal, (b) vertical, (c) helical directional fibers.
		}
		\label{fig:aniso}
	\end{figure}
}

\newcommand{\figMuscle}{
  \begin{figure}[tb]
    \centering
    \includegraphics[width=\columnwidth]{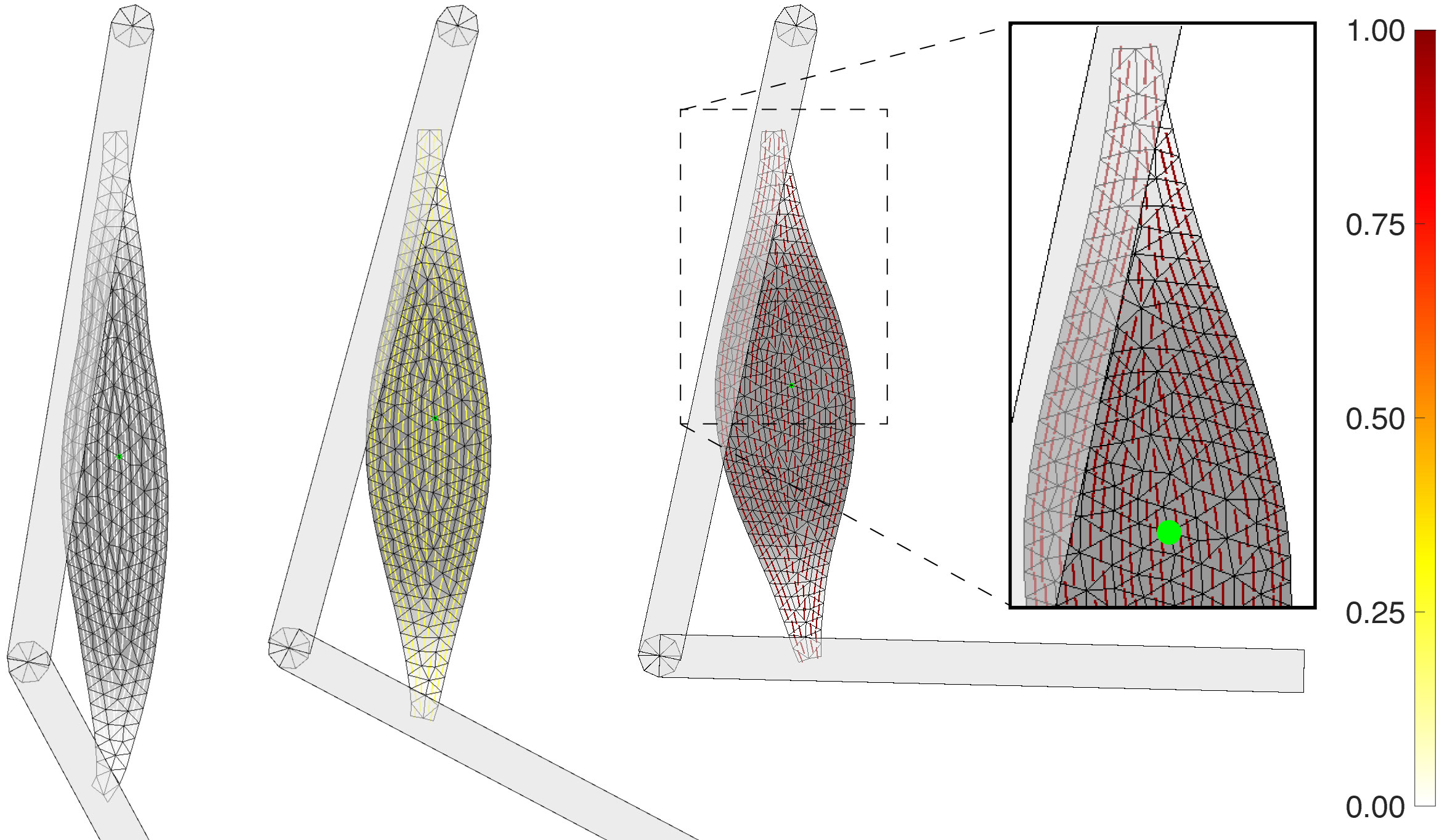}
    \caption{
      \muscle{} scene that combines \conjac{} with rigid body dynamics.
      There is one dynamic node in the middle of the muscle, colored green in the inset subfigure.
      The colored lines in the muscle foreground show the fiber directions of the anisotropic Fung material, activated from white to yellow to red.
      The muscle background is color coded in gray with the stiffness of the SNH material.
    }
    \label{fig:muscle}
  \end{figure}
}

\newcommand{\mysizeC}{1.5cm}
\newcommand{\figCut}{
	\begin{figure}[t]
		\centering
		\subcaptionbox{\label{fig:cut1}}{
			\includegraphics[height=\mysizeC]{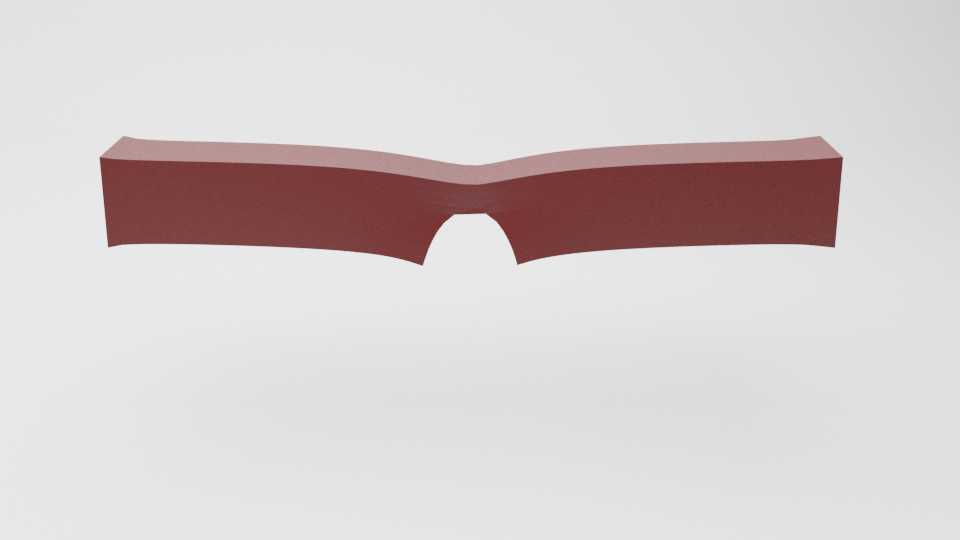}
		}
		\subcaptionbox{\label{fig:cut2}}{
			\includegraphics[height=\mysizeC]{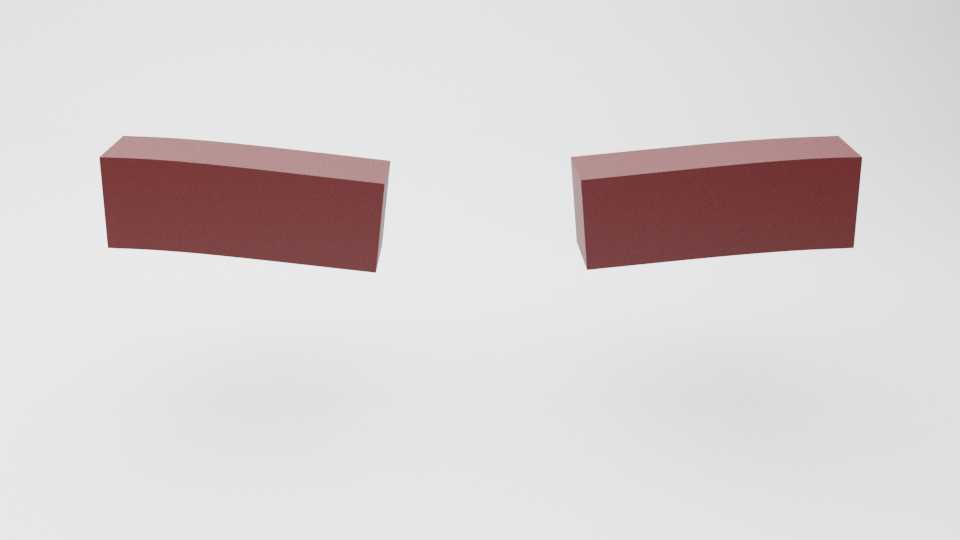}
		}
		\subcaptionbox{\label{fig:cut3}}{
			\includegraphics[height=\mysizeC]{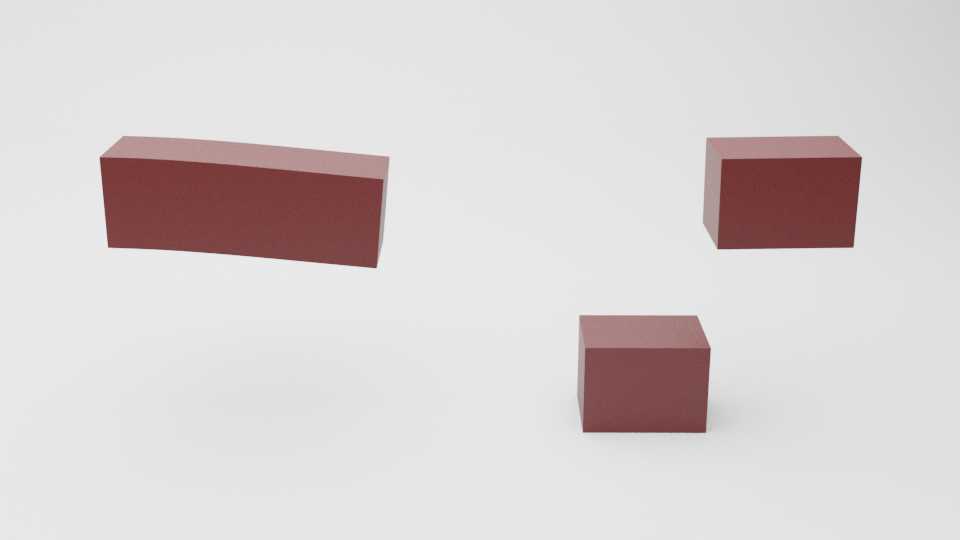}
		}
		\vspace{-0.1in}
		\caption{
			\barCut{} scene where a bar is stretched apart, (a) cut, and (b) fully separated into two halves.
			(c) The right piece is further cut into two pieces.
			The left and middle pieces are dynamic, while the right piece becomes quasistatic.
		}
		\label{fig:cut}
	\end{figure}
}

\newcommand{\mysizeA}{2.35cm}
\newcommand{\figBunny}{
	\begin{figure}[t]
		\captionsetup[subfigure]{}
		\centering
		\subcaptionbox{\label{fig:bunny1}}{
			\includegraphics[height=\mysizeA]{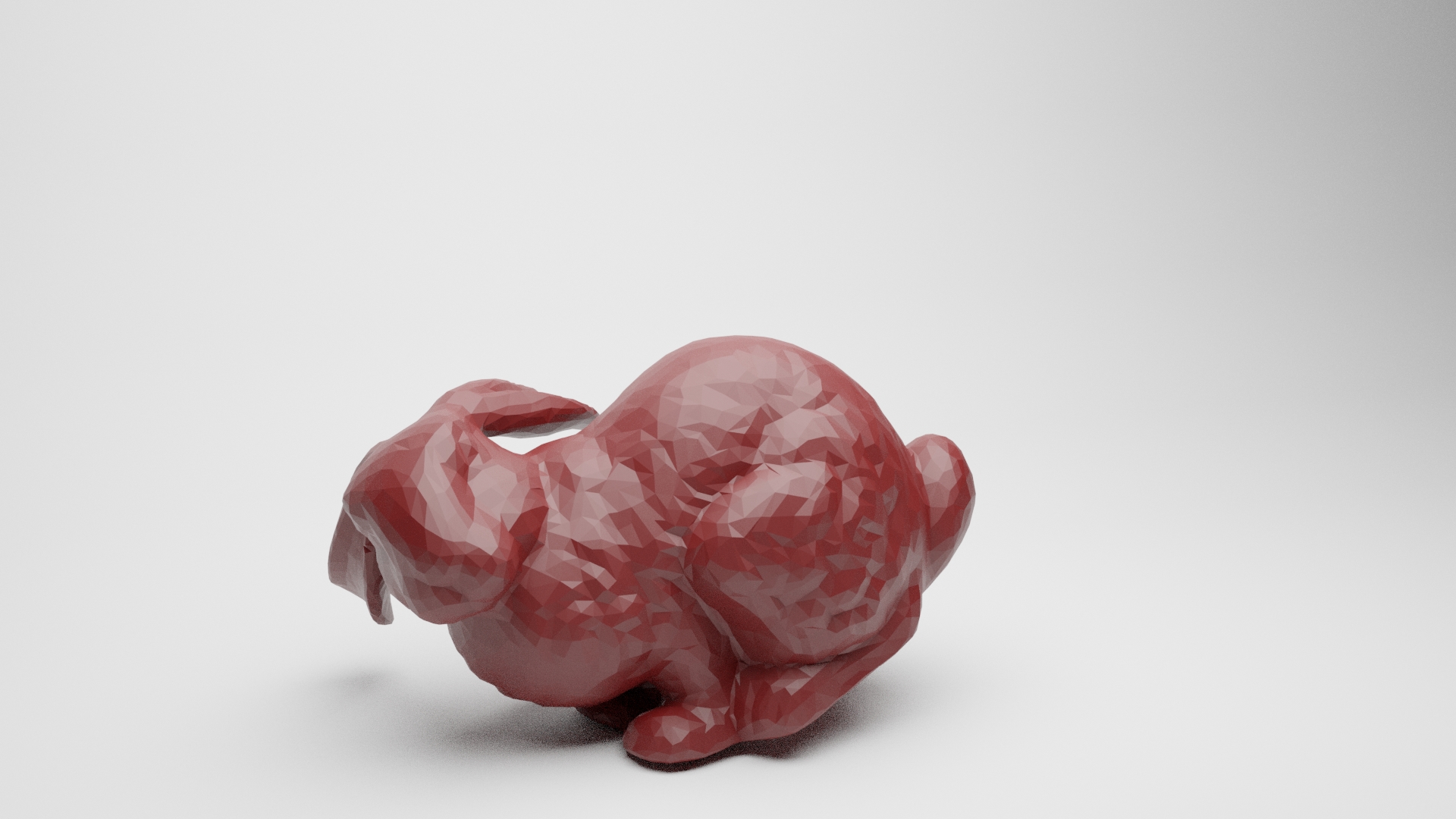}
		}
		\subcaptionbox{\label{fig:bunny2}}{
			\includegraphics[height=\mysizeA]{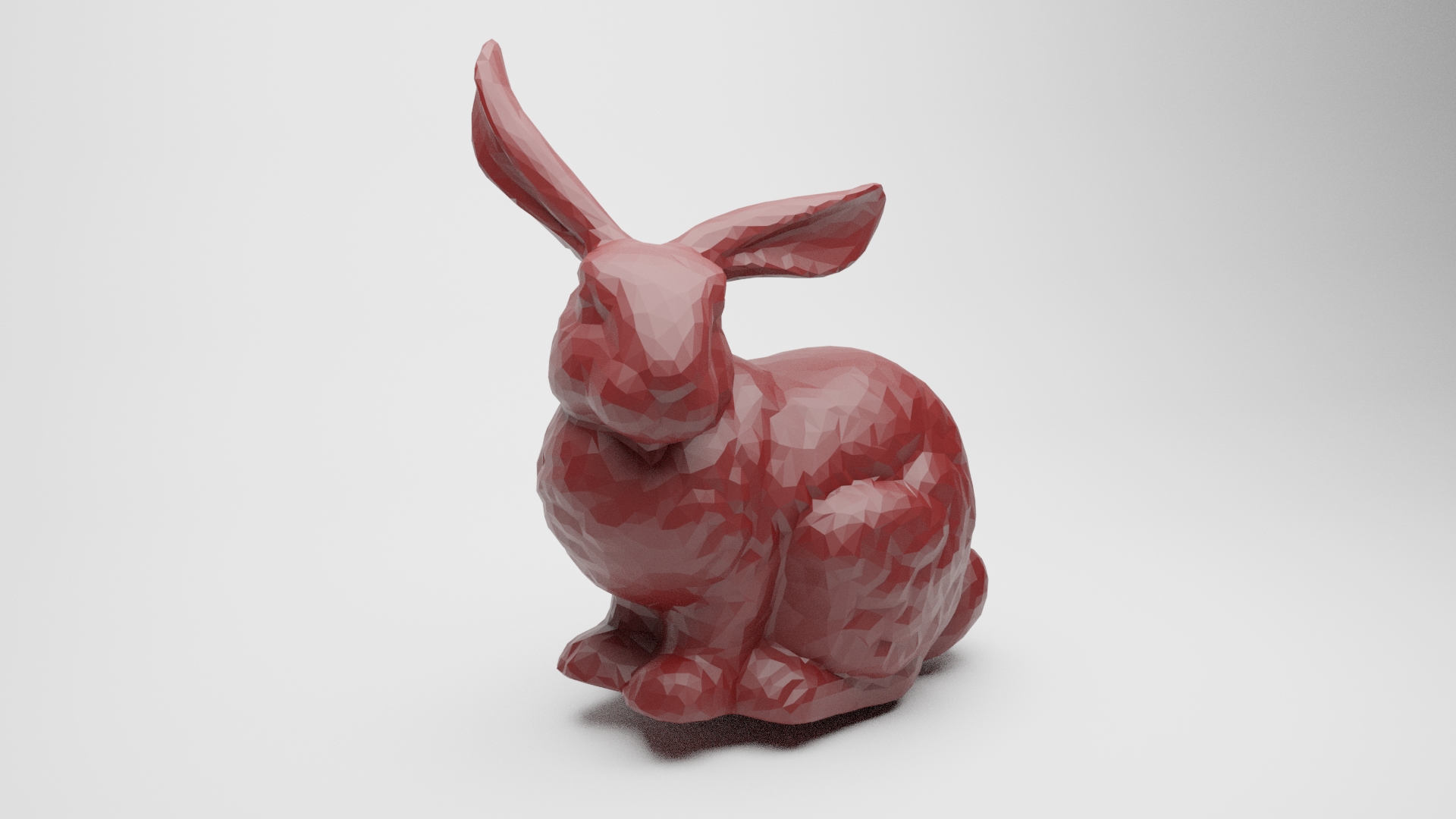}
		}
		\vspace{-0.15in}
		\caption{
			\bunny{} falling, colliding with the floor, (a) deforming from the impact, (b) bouncing back up.
		}
		\label{fig:bunny}
	\end{figure}
}

\newcommand{\mysizeD}{2.35cm}
\newcommand{\figArma}{
	\begin{figure}[t]
		\captionsetup[subfigure]{}
		\centering
		\subcaptionbox{\label{fig:arma1}}{
			\includegraphics[height=\mysizeD]{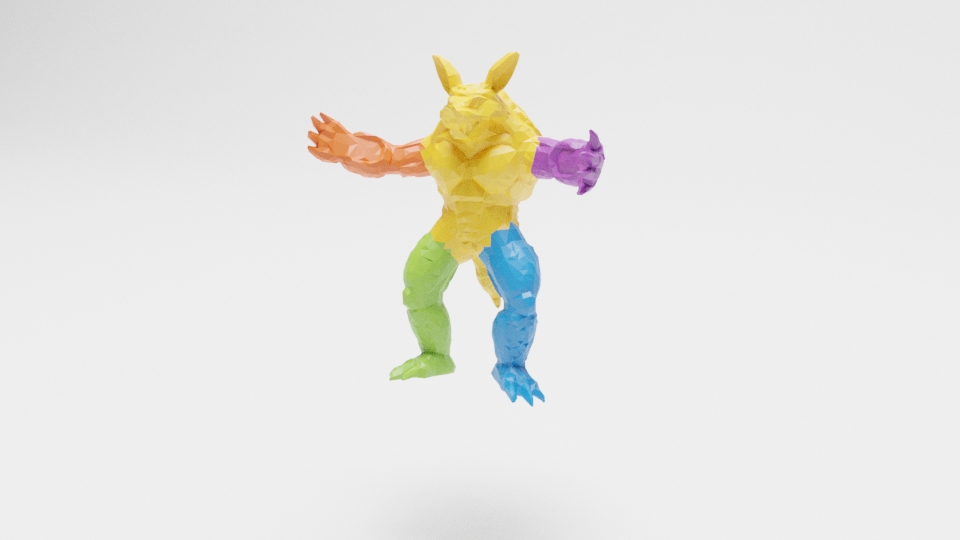}
		}
		\subcaptionbox{\label{fig:arma2}}{
			\includegraphics[height=\mysizeD]{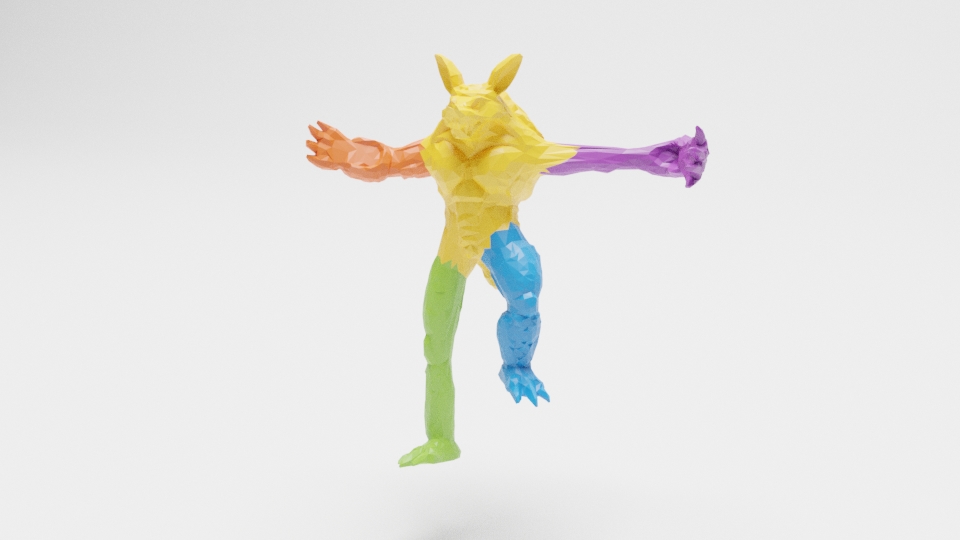}
		}
		\vspace{-0.15in}
		\caption{
			\editBegin{}
			\arma{} (a) resting under gravity and (b) having individual limbs pulled.
			\editEnd{}
		}
		\label{fig:arma}
	\end{figure}
}

\newcommand{\figSdot}{
  \begin{figure}[t]
    \centering
    \includegraphics[height=2.2in,trim={0in 0in 6.665in 2.8in},clip]{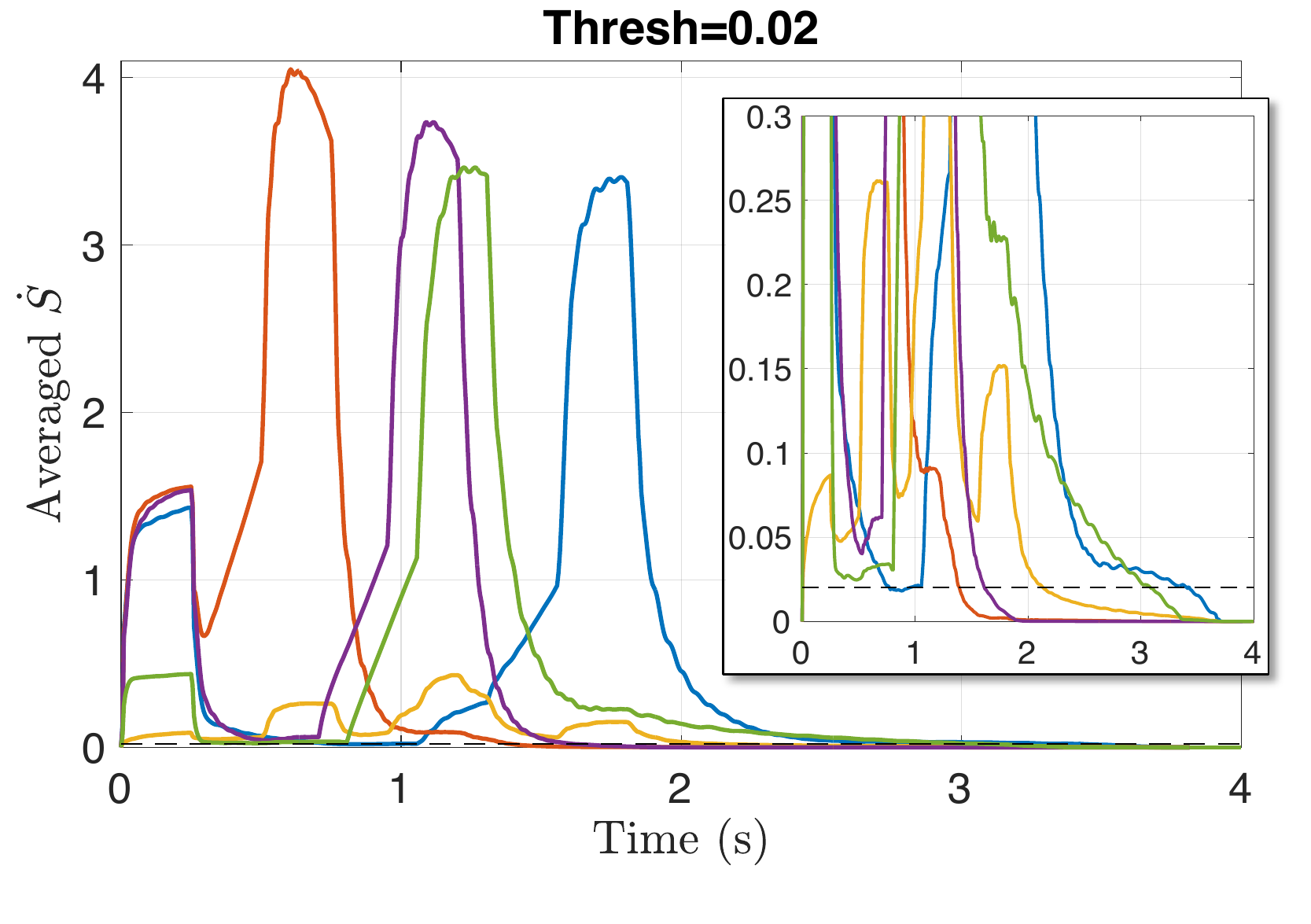}\\
    \includegraphics[height=2.2in,trim={6.665in 0in 0in 2.8in},clip]{figs/armaAdaptivity.pdf}
    \caption{
	  \editBegin{}
      Average $\dot\SS$ values of the representative nodes, color coded to correspond to the \arma{} mesh. 
      The top plot is with the threshold set to 0.02, and the bottom with 0.10.
      Zoomed versions are shown as inset figures.
      With a higher threshold, the representative nodes become quasistatic earlier.
	  \editEnd{}
    }
    \label{fig:sdot}
  \end{figure}
}

\begin{document}
%
\title{Condensation Jacobian with Adaptivity}

\author{Nicholas~J.~Weidner,
        Theodore~Kim,
         and~Shinjiro~Sueda
\IEEEcompsocitemizethanks{
\IEEEcompsocthanksitem N.J. Weidner and S. Sueda are with the Department of Computer Science \& Engineering, Texas A\&M University, College Station, TX, 77843.
\IEEEcompsocthanksitem T. Kim is with the Department of Computer Science, Yale University, New Haven, CT, 06520.
}
\thanks{Manuscript received January 1, 202X; revised January 1, 202X.}}

\IEEEtitleabstractindextext{%
\begin{abstract}


We present a new approach that allows large time steps in dynamic simulations.
Our approach, \conjac{}, is based on condensation, a technique for eliminating many degrees of freedom (DOFs) by expressing them in terms of the remaining degrees of freedom.
In this work, we choose a subset of nodes to be \textit{dynamic} nodes, and apply condensation at the velocity level by defining a linear mapping from the velocities of these chosen dynamic DOFs to the velocities of the remaining \textit{quasistatic} DOFs.
We then use this mapping to derive reduced equations of motion involving only the dynamic DOFs. 
We also derive a novel stabilization term that enables us to use complex nonlinear material models.
\conjac{} remains stable at large time steps, exhibits highly dynamic motion, and displays minimal numerical damping.
In marked contrast to subspace approaches, \conjac{} gives exactly the same configuration as the full space approach once the static state is reached.
\editBegin{} Furthermore, \conjac{} can automatically choose which parts of the object are to be simulated dynamically or quasistatically. \editEnd{}
Finally, \conjac{} works with a wide range of moderate to stiff materials, supports anisotropy and heterogeneity, handles topology changes, and can be combined with existing solvers including rigid body dynamics.



\end{abstract}

}


\maketitle

\IEEEdisplaynontitleabstractindextext

%
\IEEEpeerreviewmaketitle





Physics-based simulation of dynamic deformable objects has a long history in computer graphics.
Starting with the work by Terzopoulos et al.~\cite{Terzopoulos1987}, algorithms for physics-based animation have steadily become an integral part of the visual effects pipeline.
Over the years, various improvements have been made, including: novel energy formulations \cite{Smith2018}, inversion recovery/safety \cite{Irving2004,Kim:2019}, novel Eulerian/Largrangian formulations \cite{Levin2011,Jiang2016}, and completely new time stepping schemes \cite{Muller2007,Bouaziz2014}.

Computational efficiency is one of the most important aspects of simulation.
Real-time applications such as games and virtual surgery have strict computational budgets for physics, while
offline applications such as movies need efficiency so that artists can quickly iterate on designs.
However, efficiency comes at a price.
Various works have made dynamic simulation of deformable objects extremely efficient, but they inescapably introduce limitations.
To tackle this issue, we introduce a novel, reduced coordinate approach that has the following desirable properties:
\begin{itemize}
  \item Reproduces \textit{exactly the same static configuration} as the standard finite element (FE) approach.
  \item Supports complex nonlinear materials, including heterogeneity, anisotropy, and biomechanical soft tissues.
  \item Does not require any precomputation.
  \item Supports topology changes.
  \item Retains dynamic motion at large time steps, without suffering from excessive numerical damping.
  \item Can be combined with existing frameworks, including rigid body dynamics, into a fully two-way coupled simulation.
\end{itemize}


Existing works fail with respect to at least one of these properties.
The virtual surgery simulator of Bro-Nielsen and Cotin~\cite{Bro1996} is highly efficient and produces the same static configuration as the full FE method, which is useful for predicting the behavior of a virtual organ.
However, it only supports relatively small deformations, because only linear materials can be factorized as a precomputation.
In one of the seminal works on cloth simulation, Baraff and Witkin~\cite{Baraff1998} greatly increased the efficiency of dynamics simulations by introducing a linearly implicit integration method that allowed large time steps.
However, this approach fails to retain dynamics under large time steps due to numerical damping.
One of the most important approaches to improving efficiency is subspace dynamics \cite{Pentland1989,Choi2005,Barbic2005,An2008,Pan2015,Teng2015}.
These methods achieve massive speed ups, but sacrifice local detail because the subspace dimension must be kept at a minimum.
They also require precomputation, and cannot reproduce the same solution as FE unless a prohibitively large subspace is used.

Our approach is based on condensation \cite{Paz1989}, a technique for eliminating many degrees of freedom (DOFs) by expressing them in terms of the remaining DOFs.
With \conjac{}, short for \textit{Con}densation \textit{Jac}obian, we apply condensation at the velocity level---a significant departure from previous work \cite{Wilson1974,Paz1989,Bro1996,Gao2014,Teng2015}.
We select a ``dynamic'' subset of nodes as the true DOFs of the system, and the remaining ``quasistatic'' nodes are assumed to follow the dynamic nodes in a quasistatic fashion.\footnote{Previous works have called these ``external/internal'' or ``master/slave'' nodes.}
More specifically, \conjac{} expresses the velocities of the quasistatic nodes as a linear function of the dynamic nodes by leveraging the condition that the net force acting on each quasistatic node vanishes.
We also derive a novel stabilization term that allows \conjac{} to be used with an arbitrary material model.
Previous work was limited to linear materials.


We show that most of the important dynamics of an object are captured by simulating just a few key dynamic nodes, and the remainder can be handled quasistatically.
We simulate a bar stretching, compressing, bending, and twisting with only a few (1-4) dynamic nodes placed along the central axis.
We are also able to simulate the dynamics of a dragon being pulled in various locations, and a bunny being dropped on the floor, each with only 8 dynamic nodes.
The \conjac{} approach remains stable with large time steps because the quasistatic nodes cannot move independently, which effectively removes the small vibrations that can destabilize standard FE simulators.
With \conjac{}, a strong force suddenly applied to a node is instantaneously propagated to the dynamic nodes, eliminating the numerical wave that would force a full FE simulator to take small time steps.

\conjac{} is a method for reducing the DOFs of a system, and so it is not tied to a specific time integrator.
In this paper, we showcase the strengths of \conjac{} using the popular linearly implicit integration scheme \cite{Baraff1998}.
We show that with a linearly implicit scheme, \conjac{} is computationally inexpensive, requiring only one linear solve per time step, but does not suffer excessively from numerical damping and retains all of the advantages listed earlier in the introduction.





\section{Related Work}
\label{sec:related}

Simulation of deformable objects is a well-studied subject in computer animation, and we refer the reader to excellent existing surveys and tutorials \cite{Nealen2006,Sifakis2012}.

Our method is based on condensation, a technique from structural engineering \cite{Guyan1965,Irons1965,Wilson1974}.
Originally developed for static vibrational analysis, condensation has been extended to include dynamics \cite{Paz1989}.
With these classical condensation approaches, a global generalized eigenvalue problem is solved for the reduced modes of the structure.
In our work, we use condensation to derive a linear mapping of the velocities rather than to compute the modes.

Several previous works in computer graphics are motivated by condensation.
These methods use the stiffness matrix to couple specially chosen dynamic DOFs to the remaining quasistatic nodes.
Our work is closely related to the work by Gao et al.~\cite{Gao2014} on Steklov-Poincar\'e skinning.
They achieve impressive volumetric effects for skinning using only the surface degrees of freedom, but is limited to quasistatics and corotational elasticity.
The same authors later developed a ``macroblock'' solver for grid-based discretizations, also using a stiffness matrix reduction \cite{Mitchell2016}.
By solving the macroblocks in parallel and efficiently aggregating, they quickly compute a deformation that matches the output of a standard FE solver.
However, they again rely on linear (corotational) material that can be precomputed.
Furthermore, stiff springs are used to couple deformable objects to rigid bodies, which may reduce the time step or introduce unwanted numerical damping.

One of the most important and popular approaches to improving efficiency is subspace dynamics \cite{Pentland1989,Choi2005,Barbic2005,An2008,Li:2014,Pan2015}.
Rather than simulating the full space of vertex DOFs, dynamics are performed over a reduced set of DOFs.
To address artifacts that arise from the global support of subspace basis functions, researchers have explored domain decompositions where subspaces are computed per domain.
To stitch these domains together, Barbi\v{c} and Zhao~cite{Barbic2011} used locally aligned rigid frames, while Kim and James \cite{Kim2011} used penalty forces.
These methods can achieve massive speed ups, but sacrifice local detail because the subspace dimension must be kept at a minimum.
They also require precomputations such as modal analysis and cubature optimization, so changing object topologies are challenging.
Finally, they generally do not reproduce the full FE solution unless the subspace is prohibitively large.

Condensation has also been combined with subspace dynamics.
Traditionally, only linear materials could be used, but
Teng et al.\cite{Teng2015} efficiently performed subspace condensation at runtime, allowing nonlinear materials to also be used.
However, the overall limitations remain. The subspace must be carefully constructed, and while the condensation allows objectionable artifacts to be avoided, the final deformation does not match the full FE solution.


Recently, Xian et al.~\cite{Xian2019} introduced a multigrid-based method to solve for deformation dynamics in the full space, and achieved over 40 FPS on a mesh with over 60k vertices.
However, they inherit common limitations of multigrid methods.
Without significant extensions, it is not possible to support topological changes, complex materials (heterogeneity and anisotropy), and two-way coupling with rigid body dynamics.

Finally, a number of efficient time stepping schemes have been introduced by graphics researchers.
Recently, Li et al.~\cite{Li2019} introduced a domain-decomposed optimization method for implicit numerical time integration. 
In the past two decades, Position-Based Dynamics \cite{Muller2007}, Projective Dynamics \cite{Bouaziz2014}, and ADMM \cite{Narain2016,Zhang2019} have become popular, efficient alternatives to the standard time stepping schemes.
Although initially quite limited in terms of available materials and constraints, these methods have become quite general and flexible.
These time stepping schemes work well, but are monolithic, and would require a complete rewrite of existing formulations to make them work together.
Our work is instead based on a simple mapping of velocities, which can be incorporated into a wide range of existing explicit and implicit integrators.



\section{\conjac{} Dynamics}
\label{sec:conjac}

We begin with a high-level, didactic description of \conjac{} in action.
Imagine a vertical string discretized as a sequence of 1D nodes (\ie they can only move vertically).
We fix the top node and pick the bottom node to be the \textit{dynamic} node.
The remaining nodes in the middle are labeled as \textit{quasistatic} nodes.
If we know the material properties of the string (\eg zero rest-length springs), then by assuming that the net force on each quasistatic node remains zero, we can calculate the position and velocities of all these quasistatic nodes from the position and velocity of the single dynamic node at the bottom of the string.

In this section, we will formalize this approach by deriving the linear mapping between the quasistatic and dynamic nodes of a volumetric solid composed of an \textit{arbitrary nonlinear material}.
We will then derive equations of motion that allow us to simulate the object using only the dynamic DOFs.
The remaining nodes are simulated quasistatically, so the final resting configuration exactly matches the result of a full, non-reduced FE simulator.

\subsection{\conjac{} Mapping}
\label{sec:mapping}

Once again, we select a set of \textit{dynamic} nodes that are the exposed degrees of freedom of the system.
The remaining \textit{quasistatic} nodes move so that their net force always resolves to zero.
The \conjac{} framework uses the linear mapping that enforces this condition between the dynamic and quasistatic nodal velocities:
\begin{equation}
  \label{eq:vq}
  \vv_q = \Jqd \vv_d,
\end{equation}
where $\Jqd$ is the Jacobian term that we will derive in the rest of this section. 
Given any velocities of the dynamic nodes, $\vv_d$, this mapping allows us to compute the velocities of the quasistatic nodes, $\vv_q$.

The derivation of $\Jqd$ in \autoref{eq:vq} starts with a linearization of the forces, popularized by Baraff and Witkin~\cite{Baraff1998} and extensively used by other researchers \cite{Nealen2006}.
We approximate the implicit force at the next time step as:
\begin{equation}
  \label{eq:flinear0}
  \ff = \ff^0 + \KK^0 ( \xx - \xx^0 ),
\end{equation}
where the superscript~$0$ denotes the quantities at the current time step, and $\KK = \partial \ff / \partial \xx$ is the tangent stiffness matrix.
Substituting the next velocity as $\vv = (\xx - \xx^0)/h$, where $h$ is the step size, we obtain:
\begin{equation}
  \label{eq:flinear1}
  \ff = \ff^0 + \KK^0 h \vv.  
\end{equation}
We follow previous condensation work \cite{Wilson1974,Paz1989,Bro1996,Gao2014,Teng2015} and partition each of the terms into dynamic and quasistatic quantities:
\begin{equation}
  \label{eq:blocks}
  \begin{pmatrix}
    \ff_d\\
    \ff_q
  \end{pmatrix}
  =
  \begin{pmatrix}
    \ff_d^0\\
    \ff_q^0
  \end{pmatrix}
  +
  h
  \begin{pmatrix}
    \Kdd^0 & \Kdq^0 \\
    \Kqd^0 & \Kqq^0
  \end{pmatrix}
  \begin{pmatrix}
    \vv_d\\
    \vv_q
  \end{pmatrix}.
\end{equation}
Since we are interested in applying the zero net-force condition on the quasistatic nodes, we extract the bottom row of \autoref{eq:blocks}.
After moving $\ff_q^0$ and $h$ to the left hand side (LHS), we have:
\begin{equation}
  \label{eq:fqLinear}
  \frac{1}{h} \left( \ff_q - \ff_q^0 \right) = \Kqd^0 \vv_d + \Kqq^0 \vv_q.
\end{equation}
Our goal is to obtain zero net-force on the quasistatic nodes, so we set the force vectors to zero.
(We will return to this point in \autoref{sec:stab}.)
Rearranging \autoref{eq:fqLinear} in the form of \autoref{eq:vq}, $\vv_q = \Jqd \vv_d$, we obtain our \textit{con}densation \textit{Jac}obian (\conjac{}):
\begin{equation}
  \label{eq:Jb}
  \Jqd = -(\Kqq^0)^{-1} \Kqd^0.
\end{equation}
Moving forward, we will drop the superscript $0$ from $\KK$, with the understanding that these quantities are evaluated at the current time step.


\subsection{Equations of Motion}

Armed with the \conjac{} mapping in \autoref{eq:Jb}, we are now ready to derive the equations of motion.
First, we define an expanded mapping that includes both quasistatic and dynamic nodes:
\begin{equation}
  \label{eq:Jb2}
  \vv = \JJ \vv_d,
  \quad
  \vv =
  \begin{pmatrix}
    \vv_d \\
    \vv_q
  \end{pmatrix},
  \quad
  \JJ =
  \begin{pmatrix}
    \II \\
    \Jqd
  \end{pmatrix},
\end{equation}
where $\II$ is the identity matrix.
This mapping passes the dynamic velocities through untouched, while applying the \conjac{} mapping defined by \autoref{eq:Jb} to the quasistatic velocities.
Taking the time derivative of \autoref{eq:Jb2}, we have:
\begin{equation}
  \dot\vv = \JJ \dot\vv_d + \dot\JJ \vv_d. 
\end{equation}
Plugging $\dot\vv$ into Newton's second law, $\MM \dot\vv = \ff$, rearranging the terms, and left multiplying by $\JJ^\top$, we get:
\begin{equation}
  \label{eq:transformation}
  \JJ^\top \MM \JJ \dot\vv_d = \JJ^\top \left( \ff - \MM \dot\JJ \vv_d \right).
\end{equation}
The LHS matrix, $\JJ^\top \MM \JJ$, is the effective inertia tensor acting on the dynamic nodes.
This generalized inertia includes not only the self inertia of the dynamic nodes but also the inertia of the quasistatic nodes, since any motion of the dynamic nodes automatically causes the quasistatic nodes to move.
The right hand side (RHS) vector is pre-multiplied by the Jacobian transpose, $\JJ^\top$.
Since $\JJ^\top = \begin{pmatrix} \II & \Jqd^\top \end{pmatrix}$, the forces acting on quasistatic nodes are left-multiplied by $\Jqd^\top$ to project away the null-space.
Finally, since the goal of our approach is to approximate dynamics while preserving quasistatics, we ignore the quadratic velocity vector on the RHS involving $\dot\JJ$, which disappears when $\vv$ is zero \cite{Shabana2013}.
In our examples, the lack of the quadratic velocity vector did not cause any visual artifacts.

The \conjac{} mapping can be used with a variety of time stepping schemes.
In this work, we use the popular linearly implicit (which we call ``\vanilla{}'') formulation \cite{Baraff1998,Nealen2006,Muller2008,Lloyd2012}.
This integration scheme is easy to implement, requiring only a single linear solve per time step.
\begin{equation}
  \label{eq:vanilla}
  \left( \MM - \beta h^2 \KK \right) \vv = \MM \vv^0 + h \ff.
\end{equation}
Here, the tangent stiffness matrix, $\KK$, is evaluated at the current time step, but we have dropped the superscript for brevity.
In addition to the $h^2$ factor in the stiffness term in \autoref{eq:vanilla}, we also apply a positive factor $\beta$ to control the amount of damping \cite{Barbic2005,Xu2017}.
If we increase $\beta$, the simulation becomes more stable but at the cost of added numerical damping.

We obtain \textit{our final} \conjac{} \textit{equations of motion} by projecting \vanilla{} with the Jacobian:
\begin{equation}
  \label{eq:conjac}
    \JJ^\top \left( \MM - \beta h^2 \KK \right) \JJ \vv_d = \JJ^\top \left( \MM \vv^0 + h \ff \right).
\end{equation}
We solve this linear system at every time step for the new dynamic velocities, $\vv_d$.
Once the dynamic velocities are computed, we compute the quasistatic velocities as $\vv_q = \Jqd \vv_d$.
Then, as explained in the next section, we apply stabilization to the positions at the end of the time step.


\subsection{Stabilization}
\label{sec:stab}

The Jacobian, $\Jqd$, defined in \autoref{eq:Jb} can cause large errors for nonlinear materials, due to the linear approximation introduced in \autoref{eq:flinear0}.
Since we are applying condensation at the velocity level, after taking a time step, the quasistatic forces inevitably contain small non-zero values, which implies that the LHS of \autoref{eq:fqLinear} is not always zero.
In particular, the \textit{current} force acting on the quasistatic nodes, $\ff_q^0$, is not exactly balanced, and contains small non-zeros.
(On the other hand, the implicit force at the \textit{next} time step, $\ff_q$, is what we want to eliminate, so it is set to zero.)

This observation allows us to compute the ``residual'' velocity that drives the quasistatic nodes back to the zero net-force state.
If we do not throw away $\ff_q^0$ from \autoref{eq:fqLinear}, we obtain:
\begin{equation}
\label{eq:stab}
  \vv_q = \Jqd \vv_d + \bb_q, \quad \bb_q = -\frac{1}{h} (\Kqq^0)^{-1} \ff_q^0.
\end{equation}
This Baumgarte-like stabilization term, $\bb_q$, is the key term that makes our approach work, even in the presence of linearization artifacts \cite{Baumgarte1972}.
Rather than modifying the velocities, we apply this stabilization term when we update the positions.
We multiply this factor by a scalar parameter $\gamma$ that controls the strength of the stabilization.
The position updates for dynamic and quasistatic nodes are then:
\begin{equation}
  \label{eq:xstab}
  \begin{split}
    \xx_d &= \xx_d^0 + h \vv_d\\
    \xx_q &= \xx_q^0 + h (\vv_q + \gamma \bb_q).
  \end{split}
\end{equation}

When applied to the position, this stabilization term becomes exactly a Newton correction term: $\Delta \xx = -\gamma \Kqq^{-1} \ff_q$.
In other words, we apply one scaled Newton step at the position level after taking a velocity step, with $\gamma=1$ corresponding to a full Newton step.
In practice, we found that a full Newton step can sometimes cause instabilities.
The best value can be obtained with a line search, but we found that simply setting $\gamma=1/3$ worked well for our examples (unless otherwise stated).

Without the stabilization term $\bb_q$, the object becomes visibly distorted due to the accumulation of error, and can eventually blow up.
This term had not been derived in previous approaches because linearization does not cause any drift in linear materials.
This stabilization approach is both effective and efficient.
An alternative approach based on pre- or post-stabilization may work as well \cite{Cline2003,Weinstein2006}, but we speculate that they will be less efficient and more difficult to implement.

\subsection{Time Stepping}

\begin{algorithm}[t]
 \caption{
 \conjac{} pseudocode
 }
 \label{alg:conjac}
 \begin{algorithmic}[1]
  \State (Optional) Initialize the quasistatic positions
  \State Compute $\MM$
  \While{simulating}
    \State Compute $\ff, \KK$
    \State Compute $\JJ$, $\bb$
    \State Solve for $\vv_d$ (\autoref{eq:conjac})
    \State Compute $\vv_q = \Jqd \vv_d$
    \State Update $\xx$ (\autoref{eq:xstab})
  \EndWhile
 \end{algorithmic}
\end{algorithm}

\begin{algorithm}[t]
 \caption{
 \vanilla{} pseudocode
 }
 \label{alg:vanilla}
 \begin{algorithmic}[1]
  \State Compute $\MM$
  \While{simulating}
    \State Compute $\ff, \KK$
    \State Solve for $\vv$ (\autoref{eq:vanilla})
    \State Update $\xx = \xx^0 + h \vv$
  \EndWhile
 \end{algorithmic}
\end{algorithm}

The overall simulation pseudocode for \conjac{} using linearly implicit Euler \cite{Baraff1998} is shown in \autoref{alg:conjac}.
For comparison, we also show the \vanilla{} pseudocode, also using linearly implicit integration, in \autoref{alg:vanilla}.


With \vanilla{}, the performance bottleneck is the linear solve for the new velocities (line 4).
On the other hand, with \conjac{}, solving for the new dynamic velocities is not the bottleneck because \autoref{eq:conjac} is small.
Instead, the bottleneck is in forming the Jacobian (line 5), which involves a series of solves by $\Kqq$, which cannot be prefactored for nonlinear materials.

With our current implementation, each time step of \conjac{} (lines 4-8 in \autoref{alg:conjac}) is about 20\% slower than a time step of \vanilla{} (lines 3-5 in \autoref{alg:vanilla}) with 4 dynamic nodes, and 40\% slower with 10 dynamic nodes (see \autoref{fig:ecost}).
However, we more than make up for this difference because \conjac{} allows much bigger time steps for the same amount of dynamic behavior.



The initial nonlinear solve for the quasistatic positions in \conjac{} (line 1 in \autoref{alg:conjac}) can be costly, but it only needs to be performed once at the beginning of the simulation.
We do not need to run this expensive nonlinear optimization within the simulation loop because of the stabilization term from \autoref{sec:stab}.
In fact, it is even possible to skip the initial nonlinear solve, since the stabilization term eventually eliminates the drift and drives quasistatic nodes to their zero net-force state over time.

%
%
%

\editBegin{}

\section{Adaptivity}

The liveliness of a \conjac{} simulation is tied to the number of dynamic nodes in the scene.
We can choose to place dynamic nodes only in regions where dynamics are desired to avoid unnecessarily increasing the bottleneck.
To generalize objects so that they are still lively and optimized in novel deformations and environments, we introduce a concept of adaptivity---we turn on/off the dynamic nodes \textit{at runtime}.
We assume that we know \textit{a priori} a subset of mesh nodes that can become dynamic, which we call the ``representative'' nodes.
As we show in \autoref{sec:results}, this number does not need to be very high to get rich deformations.
For example, in the \arma{} mesh shown in \autoref{fig:arma}, this subset consists of 5 representative nodes, placed in the extremities of the four limbs and in the center of the torso.
During runtime, we automatically decide which of these representative nodes should be dynamic or quasistatic, depending on our novel ``liveliness'' metric.
This cuts down on unnecessary solves which speeds up simulations, and improves the robustness of scenes.
In the rest of this section, we will describe our liveliness metric (\autoref{sec:metric}) and then discuss the necessary changes to the \conjac{} algorithm to minimize expensive matrix resizing and slicing operations that occur when dynamic nodes are turned on and off at runtime (\autoref{sec:matrix}).

\subsection{Adaptivity Metric}
\label{sec:metric}

To quantify the liveliness of a node, we want a metric that captures how a local region of the mesh is deforming differently from its neighborhood regions.
We are interested in capturing the differences in the rate of change of deformation.
Therefore, rather than using the deformation gradient $\FF$, we use the time derivative of the deformation gradient $\dot\FF$.
In particular, we look at the average change in stretching speed over a local group of tetrahedral elements. 
Stretch is a very insightful local measurement into how much our object is actually deformed rather than undergoing rigid motion, and change in stretch captures activity instead of a deformed settled state.

To derive this change in stretch over time, or $\dot\SS$, we look at the deformation gradient $\FF$ based on our material matrix $\DD_m$ and spacial matrix $\DD_s$:
\begin{equation}
  \label{eq:defgrad}
  \begin{split}
  \DD_m
  &=
  \begin{pmatrix}
     \overline\xx_1 - \overline\xx_0 | \overline\xx_2 - \overline\xx_0 | \overline\xx_3 - \overline\xx_0
  \end{pmatrix}\\
    \DD_s
  &=
  \begin{pmatrix}
     \xx_1 - \xx_0 | \xx_2 - \xx_0 | \xx_3 - \xx_0
  \end{pmatrix}\\
  \FF
  &=
  \DD_s \DD_m^{-1}.
  \end{split}
\end{equation}
These matrices are a formulation of our nodal material positions in world space, $\overline\xx$, and our current time step's deformed nodal positions in world space, $\xx$, respectively.
$\FF$ can also be decomposed into rotation and stretch components using the polar decomposition:
\begin{equation}
  \label{eq:fdecomp}
  \FF = \RR \SS.
\end{equation} 
Substituting our deformed positions for velocities allows us to instead formulate a velocity gradient.
Using the chain rule, this velocity gradient can be similarly decomposed just like $\FF$, using nodal velocities $\vv_i$ instead of positions $\xx_i$ \cite{SanchezBanderas2018}:
\begin{equation}
  \label{eq:velgrad}
  \begin{split}
  \dot\DD_s
  &=
  \begin{pmatrix}
     \vv_1 - \vv_0 | \vv_2 - \vv_0 | \vv_3 - \vv_0
  \end{pmatrix}\\
  \dot\FF
  &=
  \dot\DD_s \DD_m^{-1}\\
  \dot\FF
  &=
  \dot\RR \SS + \RR \dot\SS.
  \end{split}
\end{equation} 
From this decomposition we can rearrange and solve for $\dot\SS$:
\begin{equation}
  \label{eq:sdot}
  \dot\SS
  =
  \RR^\top \left(\dot\FF - \dot\RR \SS \right).
\end{equation}
A singular value decomposition of $\FF$ gives us definitions for $\RR$ and $\SS$:
\begin{equation}
  \label{eq:fsvd}
  \quad
  \FF = \UU \Sigma \VV^\top
  \quad
  \RR = \UU \VV^\top
  \quad
  \SS = \VV \Sigma \VV^\top.
\end{equation}
The slightly more complicated piece we still need is $\dot\RR$, which we can decompose as follows:
\begin{equation}
  \dot\RR = \frac{\partial\RR}{\partial\FF} \colon \dot\FF.
\end{equation}
We compute $\partial\RR / \partial\FF$ in closed form by following the work of Smith et al.~\cite{Smith:2019:AEI}.
(The pseudocode is given in \autoref{sec:pseudo_rotation}.)
Computationally speaking, when using a material model such as the Stable Neo-Hookean material~\cite{Smith2018}, the expensive SVD component of these operations is already required so the only additional work needed for this new metric is the relatively inexpensive $\dot\RR$ value.

Once this metric is defined per tetrahedron, we want to quantify this measurement for each representative node so we know whether a particular representative node should be dynamic or quasistatic at a given time step.
$\dot\SS$ is a a matrix whose coefficients represent the speed of change of the deformation, so by using the absolute value of these coefficients to ignore direction and the average of them to alleviate outliers, we arrive at a scalar value giving us a good idea how much deformation is taking place.
In other words, the liveliness measure of the $j^{th}$ representative node is:
\begin{equation}
	\text{metric}_j = \text{mean} \left( \left[ \text{vec}(|\dot\SS_1|)^\top \; \cdots \; \text{vec}(|\dot\SS_m|)^\top \right] \right),
\end{equation}
where $m$ is the number of tetrahedra in the region owned by the $j^{th}$ representative node.
(The pseudocode is given in \autoref{sec:pseudo_metric}; in our actual implementation, we use a weighted average using the volume of each element.)
To further account for potential noise in the metric, we expand this by averaging these metrics across a window of past time steps.
Our final scalar value is compared against a threshold of desirable motion and the end result is a dynamic node that can gracefully revert to a quasistatic state when the local deformations around it are not worth spending the increased number of solves to capture.
In the extreme case, when the dynamic motion has mostly died down, the simulation is driven entirely from the stabilization term, with all nodes moving in a quasistatic fashion.

\subsection{Sparse Matrix Handling}
\label{sec:matrix}

\begin{algorithm}[t]
 \editBegin{}
 \caption{
 \conjac{} with adaptivity
 }
 \label{alg:conjacdyn}
 \begin{algorithmic}[1]
  \State (Optional) Initialize the quasistatic positions
  \State Compute $\MM$
  \While{simulating}
    \State Compute $\ff, \KK, \dot\SS$
    \For{representative nodes}
      \State Compute metric
      \State Set as dynamic or quasistatic
    \EndFor
    \State Adjust sparse values of $\KK$ and $\Kqd$
    \State Compute $\JJ$, $\bb$
    \State Solve for $\vv_d$ (\autoref{eq:conjac})
    \State Compute $\vv_q = \Jqd \vv_d$
    \State Update $\xx$ (\autoref{eq:xstab})
  \EndWhile
 \end{algorithmic}
\end{algorithm}

\figDragon

Introducing this mechanic for flipping the state of a subset of our nodes forces us to take another look at our \conjac{} algorithm.
The construction of our Jacobian is reliant on both the $\Kqq$ and $\Kqd$ matrices, which can now vary in size \textit{at runtime} depending on the number of dynamic and quasistatic nodes.
Because the construction of our Jacobian is our bottleneck, we want to ensure that we are not introducing new overhead on top of the existing \conjac{} algorithm that used fixed-sized matrices.
More specifically, some operations such as large matrix allocations and sparsity pattern analyses (row/column permutations and symbolic analysis) that usually take place in the simulation setup phase now must happen each time step that a representative node changes from dynamic to quasistatic and vice versa. 

Predefining the subset of representative nodes that can flip between states has a major advantage in combatting this issue.
Without adaptivity, we used \autoref{eq:Jb}, which required the entire stiffness matrix, $\KK$, to be pre-partitioned into $\Kqq$ and $\Kqd$.
(Since the number of quasistatic nodes is much larger than the number of dynamic nodes, $\Kqq$ is almost the same size as $\KK$, but $\Kqd$ is a tall and skinny matrix.)
With adaptivity, rather than partitioning $\KK$ into $\Kqq$ and $\Kqd$ at runtime, we instead adjust the non-zeros of $\KK$ and $\Kqd$ on the fly to account for the changing number of dynamic nodes.
Specifically: (1) we zero out the row and column of $\KK$ corresponding to each of the dynamic nodes and place a negative one on the diagonal; and (2) we replace the rows of $\Kqd$ corresponding to the dynamic nodes with the identity matrix.
In this context, to zero-out refers to explicitly setting a sparse value to zero rather than to change the sparsity.
(The pseudocode for these operations is provided in \autoref{sec:pseudo_adjusted}.)
Calling these adjusted matrices $\KK_A$ and $\KK_{qdA}$, the Jacobian can be computed as:
\begin{equation}
  \label{eq:JA}
  \JJ = -\KK_A^{-1} \KK_{qdA},
\end{equation}
instead of Eqs.~\ref{eq:Jb} and \ref{eq:Jb2}.
This allows us to take advantage of the reduced number of solves without shifting around, reallocating, or reanalyzing unnecessary data in the sparse matrices.

\editEnd{}


\section{Results}
\label{sec:results}

We implemented our system in MATLAB and ran the simulations on a consumer laptop with an Intel Core i9-9880H CPU @ 2.3 GHz and 16 GB of RAM.
We use MEX for filling the force vector and the stiffness matrix, and CHOLMOD for sparse linear factorizations and solves \cite{Davis2006}.
The scene parameters are listed in \autoref{tab:params}.
All of the objects are table-top sized---roughly 5-15 cm across, weighing a few hundred grams.
For all results, we use the Stable Neo-Hookean (\SNH{}) base material \cite{Smith2018}.
This material is stable under inversion, but like any non-linear material, it can still need a Newton solve plus line search to maintain stability under large deformations.
We found that when used with a linearly implicit scheme, it must be heavily damped when using a large time step, especially when Poisson's ratio, $\nu$, is close to 0.5.

\begin{table}[b]
\center
\caption{
	List of scene parameters.
	\#vert: number of total vertices.
	\#dyn: number of dynamic vertices.
	\#elem: number of elements.
	mat: material model.
	Y: Young's modulus (Pa).
	$\nu$: Poisson's ratio.
}
\vspace{-2mm}
\label{tab:params}
\begin{tabular}{lrrrcrr}
\rowcolor{WHITE} Scene    & \#vert  & \#dyn & \#elem &    mat      &      Y      & $\nu$ \\
\cmidrule(lr){1-1} \cmidrule(lr){2-7}
\rowcolor{WHITE} \dragon  & 10456   & 0-10  & 37565  &  \SNH   & \Epos{3}{4} &  0.49 \\
\rowcolor{GRAY}  \twist   &  1029   & 1-32  &  4320  &  \SNH   & \Epos{1}{4} &  0.40 \\
\rowcolor{WHITE} \hetero  &  5915   &    1  & 29376  &  \SNH   & \Epos{1}{4} &  0.40 \\
\rowcolor{GRAY}  \aniso   &  6591   &    1  & 32832  & +\aSTVK & \Epos{1}{4} &  0.40 \\
\rowcolor{WHITE} \muscle  &   262   &    5  &   438  & +\aFUNG & \Epos{3}{4} &  0.49 \\
\rowcolor{GRAY}  \barCut  &  6050   &    2  & 29400  &  \SNH   & \Epos{1}{4} &  0.40 \\
\rowcolor{WHITE} \bunny   &  5988   &    8  & 27695  &  \SNH   & \Epos{4}{4} &  0.45 \\
\rowcolor{GRAY} \editBegin \arma \editEnd  &  5159   &    5  & 18448  &  \SNH   & \Epos{6}{4} &  0.49 \\
\end{tabular}
\end{table}


\figExtCost

\figTwist

\textbf{\dragon{}}:
We start with a 10k node dragon, shown in \autoref{fig:teaser}.
This example shows that \conjac{} presents an attractive option for efficiently producing lively simulations.
We grab the jaws and the body of the dragon and pull them in different directions.
After some time has passed, we let go, instantaneously releasing the built-up energy.
We compare the results using \conjac{} and \vanilla{}, both with time step $h\!=\,$\Eneg{5}{3} for this 1 second simulation.
For the damping factor, we use $\beta=0.5$ for \conjac{} and $\beta=3.7$ for \vanilla{} (\autoref{eq:vanilla} and \autoref{eq:conjac}).
These values were chosen by manually searching for the smallest $\beta$ values in $0.1$ increments that produced stable simulations.
As can be seen in the supplemental video, the discrepancy in the $\beta$ values are visibly significant.
\textit{Using the same $h$, \conjac{} produces highly dynamic results, whereas \vanilla{} produces heavily damped results.}
Since we are using the linearly implicit integrator, more dynamic results can be generated with \vanilla{} by reducing $h$, but this adds computational cost.
\conjac{}, on the other hand, allows large time steps while retaining interesting dynamics.
If we reduce the time step to $h\!=\,$\Eneg{2}{3} with \vanilla{}, the qualitative behavior of the dragon becomes nearly as lively as \conjac{}, but the wallclock simulation time increases to more than double the time of \conjac{} with 6 dynamic nodes.
For didactic purposes, we also include a \conjac{} simulation with 0 dynamic nodes, which produces a quasistatic simulation driven solely by the stabilization term, $\bb_q$ from \autoref{eq:stab}.
For this example, we used the stabilization factor $\gamma=1/5$, since the Newton displacements immediately after releasing the jaws and the body are extremely large.
Once we add dynamic nodes, the behavior becomes very lively, even with only 2 nodes.
The wallclock times of \conjac{} is compared to \vanilla{} in \autoref{fig:ecost}.
Virtually all of the added cost is in the triangular solves---since we require $3n_d + 1$ solves, where $n_d$ is the number of dynamic nodes, the cost increases linearly in $n_d$.
(The +1 is for computing the stabilization term, $\bb_q$.)
For most objects, 4 to 8 dynamic nodes are enough to produce convincingly dynamic results.
We discuss potential ways to improve performance in \autoref{sec:limitations}.

\figTwistInset

\textbf{\twist{}}:
Here, we show the deformation behavior of \conjac{} as we increase $n_d$, the number of dynamic nodes.
For this scene, we use \conjac{} to simulate a bar with one of its ends moved kinematically to compress, stretch, bend, and twist the bar as shown in \autoref{fig:exts}.
For $n_d = \{1, 2, 4\}$, we place the dynamic nodes at equal intervals along the central horizontal axis.
For $n_d = \{8, 16, 32\}$, we slice the bar orthogonal to the central axis at equal intervals and place 4 dynamic nodes at the corners of each of these vertical slices.
Interestingly, it becomes difficult to visually distinguish between these cases---even with 1 dynamic node, the dynamic motion is convincing.
When the dynamic nodes are placed along the central axis ($n_d=\{1,2,4\}$), we get the added ``feature'': the twisting waves are propagated instantaneously along the bar, increasing the stability of the system.
If the dynamic nodes are placed along vertical slices ($n_d=\{8,16,32\}$), we recover the twisting dynamics.



\figHetero

\textbf{\hetero{}}:
We show that \conjac{} efficiently and effectively handles heterogeneous materials.
In this example, we use \conjac{} to simulate a vertical bar with alternating layers of stiffnesses.
\autoref{fig:hetero} shows that even with only one dynamic node, we can capture the bulging of the soft layers.
Because of gravity, the lower soft layer bulges out more than the upper soft layer, even though they have the same stiffness.
Once the object reaches its static state, the final shape is exactly the same as the one generated by  \vanilla{}.



\figAniso

\textbf{\aniso{}}:
We show the effect of anisotropic materials.
On top of the base \SNH{} material, we add an anisotropic Saint Venant–Kirchhoff material (\aSTVK{}) \cite{Kim:2019}.
In this example, we use \conjac{} with one dynamic node to simulate a vertical bar with different anisotropic directions: vertical, horizontal, diagonal, and helical.
\autoref{fig:aniso} shows that when gravity compresses the bar, it deforms differently depending on the fiber directions.
Interestingly, the helical fibers induce a twisting motion.


\figMuscle

\textbf{\muscle{}}:
\conjac{} can easily be combined with existing rigid body dynamics to model a musculoskeletal system (\autoref{fig:muscle}).
In this 2D example, we combine \conjac{} with a reduced coordinate articulated rigid body framework \cite{Wang2019}.
To attach the origin and insertion nodes to the bones, we use a Jacobian mapping that expresses the velocity of these nodes as a function of the velocities of the joints.
This allows us to solve for the velocities of the muscles and joints simultaneously to give us full two-way coupling between muscles and bones, which is important because the muscle weighs more than the bones.
We use \SNH{} for the background isotropic material, and anisotropic Fung (\aFUNG{}) for the muscle fiber material \cite{Fung:2013}.
We also take advantage of \conjac's support for heterogeneity---the stiffness of the background \SNH{} material is modulated so that it is stiffer in the tendon regions than in the muscle region.
In the resulting simulation, the dynamics of the muscle is fully accounted for by a single, central dynamic node.
In total, the system is only 4-dimensional: 2 DOFs for the joints and 2 for the muscle.
Unlike quasistatic muscle simulators that assume both bones and muscles are quasistatic, with \conjac{}, we can keep the bones fully dynamic and choose how dynamic we want the muscles to be.


\figCut

\textbf{\barCut{}}:
In this example, we show that \conjac{} supports topology changes.
We start with a horizontal bar fixed at its two ends, and we cut the bar in two locations (see \autoref{fig:cut}).
We place two dynamic nodes on either side of the initial cut.
Because \conjac{} requires no precomputation, the cut can be placed anywhere.
After the second cut, the right-most piece loses all dynamic nodes and gracefully degrades into a purely quasistatic model.


\figBunny

\textbf{\bunny{}}:
In this example, we show that \conjac{} can be extended to handle frictional contact.
We drop a bunny with 8 dynamic nodes onto the floor with various starting orientations.
We follow the formulation by McAdams et al.~\cite{McAdams2011} for the contact penalty force: $\ff = K \left((1 - \alpha) \nn \nn^\top + \alpha \II \right) (\xx - \xx_s)$, where $K$ is a stiffness constant, $\nn$ is the collision normal, and $\xx_s$ is the closest point on the collision surface.
When $\alpha=0$, the spring acts only along the normal direction, and when $\alpha=1$, the spring acts isotropically.
In our experiments, we use $\alpha=0.1$.
For friction, we use the velocity filter approach by Bridson et al.~\cite{Bridson2002} to compute the post-friction velocity, $\vv^f$, of all nodes.
For the coefficient of friction, we use a global value of $\mu=0.3$.
We then use weighted least squares to compute our new dynamic velocity: $\vv_d^* = \text{argmin} \| \vv^f - \JJ \vv_d \|_{\tilde\MM}^2$, where $\tilde\MM = \MM - \beta h^2 \KK$ with $\beta=0.5$ as in other examples.
This solve is inexpensive, since we solve only for the dynamic nodes of domains in contact.
When collisions occur with quasistatic nodes, the contact information is added to the global stiffness matrix, making \conjac{} be collision-aware.
\conjac{} intelligently transfers the masses of the quasistatic nodes to the dynamic nodes, giving us a small ($24 \times 24$ in this case since there are 8 dynamic nodes) and stable system to solve at each time step.
Even with only 8 dynamic nodes, \conjac{} gives remarkably rich deformations.
For example, although the front feet and the two ears only have one dynamic node each, they undergo significant local nonlinear deformations upon contact, as shown in \autoref{fig:bunny} and the supplemental video.

\editBegin{}

\figArma

\figSdot

\textbf{\arma{}}:
In our last example, we showcase \conjac{} with adaptivity in place.
An armadillo is fixed in place by a subset of internal nodes in the center of its body.
5 representative nodes are placed in the center of 5 tetrahedral regions shown by the different colors on the mesh in \autoref{fig:arma}.
The simulation begins with all representative nodes in a deactivated (quasistatic) state before gravity introduces the initial dynamic motion that is then followed by a series of pulling forces.
As the limbs are pulled and released, the representative nodes activate and deactivate, based on the relative motion in the region.
While the body is not pulled itself, it picks many of the shockwave motions that activate it briefly multiple times.
This can be seen clearly by the motion in the tail and nose.
Even though these regions are not pulled themselves, they react realistically in a quasistatic fashion when nearby limbs are pulled.
\autoref{fig:sdot} shows the plots of our liveliness metrics.
The top figure shows the plot of the metric over time with a threshold of $0.02/s$, and the bottom with a threshold of $0.1/s$.
In other words, a representative node is dynamic as long as the average stretching speed is greater than 2\% or 10\% per second.
The inset figures show close-ups of these plots.
It can be clearly seen that with the lower threshold, the dynamics is retained longer, since the representative nodes remain active for longer.

\editEnd{}

\section{Conclusion}
\label{sec:conclusion}

\conjac{} is a new reduced coordinate approach based on condensation.
Unlike previous work, we apply condensation at the velocity level by defining a mapping that expresses the velocities of quasistatic DOFs as a linear function of the dynamic DOFs.
Compared to \vanilla{} (the standard, full FE solution), \conjac{} remains stable at large time steps and exhibits highly dynamic motion with less numerical damping.
Furthermore, \conjac{} gives the exact same configuration as \vanilla{} once the static state is reached.
To demonstrate \conjac{}'s versatility, we have shown examples involving: a wide range of materials, anisotropy and heterogeneity, topology changes, integration with rigid body dynamics, \editBegin and adaptivity. \editEnd

\subsection{Limitations \& Future Work}
\label{sec:limitations}

For \conjac{} to maintain its advantages over \vanilla{}, the dynamic nodes must not be too close to each other.
In our \dragon{} and \bunny{} examples, we manually placed the first few dynamic nodes in strategic locations (\eg dragon jaws, bunny ears), and the rest were generated randomly.
If two dynamic nodes were generated too close to each other, we reran the random generator with a different seed.

Although the stabilization term, $\bb_q$ in \autoref{eq:stab}, works well to fight the drift due to the linearization artifacts of the Jacobian, it still cannot maintain the zero net-force state on the quasistatic nodes during motion, causing visual artifacts especially when the motion is large.
Rather than taking a single Newton step, taking multiple steps would produce better results when time steps are large.
A quasi-Newton approach, where only the force vector, and not the stiffness matrix, is updated every step may yield a good balance between convergence and performance.

In our current implementation, we explicitly form $\Jqd$, which requires $3n_d$ solves with $\Kqq$, where $n_d$ is the number of dynamic nodes.
When $n_d$ is small, the bottleneck is the factorization of $\Kqq$, making \conjac{} and \vanilla{} nearly equivalent in terms of computational cost.
As we increase $n_d$, the solves start to become the bottleneck, making \conjac{} more and more expensive compared to \vanilla{}.
However, as shown in \autoref{sec:results}, \conjac{} retains important dynamics even with few dynamic nodes.
An exciting avenue of future work is to follow the work of Mitchell et al.~\cite{Mitchell2016} to decompose the object into domains, which would allow \conjac{} to scale up to a very large mesh, since then the factorizations of $\Kqq$ can be computed per-domain.
However, obtaining good multi-threaded performance would still be a major challenge, requiring careful tuning of domain sizes and topology.

Scaling \conjac{} to very large meshes would require an iterative approach, since the factorization of $\Kqq$ may not fit into memory.
This is non-trivial for the same reason above---the number of RHS vectors is $3n_d$ where $n_d$ is the number of dynamic nodes.
One approach to resolve this issue is the block Krylov method \cite{OLeary1980}, which allows the solver to share information across multiple RHS.
However, we would still need to limit $n_d$ to be relatively small to remain competitive.




An important limitation is that frictional impulses acting on quasistatic nodes cannot be accurately handled, since these nodes are not DOFs, and so their frictional impulses can only be satisfied in a least squares sense.
This is, however, a limitation common to all reduced coordinate approaches.
Therefore, our approach is most suitable when the effects of friction are not too large.

We have found experimentally that \conjac{} does not work well for very soft objects, due to severe linearization artifacts.
For similar reasons, \conjac{} cannot handle extremely fast rotational motion.
For these types of simulations, we may need to run Newton's method to convergence, rather than using the linearly implicit Euler scheme.

\conjac{} can suffer from locking artifacts with hard constraints if these constraints are applied to quasistatic nodes.
In such cases, an averaged or softened constraint will need to be applied, or new dynamic nodes must be inserted \cite{Bergou2007,Tournier2015,Andrews2017}.

Finally, we are interested in exploring adaptive time step integrators, such as Runge-Kutta-Fehlberg or MATLAB's \texttt{ode45} \cite{Fehlberg1969,Shampine1997}. 
Given \conjac{}'s stability at large time steps even with an explicit integrator, these adaptive methods have the potential to reduce the number of total time steps substantially.
\editBegin{} Combining temporal adaptivity with spatial adaptivity presented in the paper could enable highly lively animations at a low wall-clock cost. \editEnd{}

\appendices

\editBegin{}

\section{Miscellaneous Pseudocode}

\subsection{Rotation Derivative}
\label{sec:pseudo_rotation}

Given $\FF$ and $\dot\FF$, we compute $\dot\RR$ with the following function:

\begin{verbatim}
function [A] = Rdot(F, Fdot)
  Rgradient = DRDF(F); % Smith et al. 2019
  fdot = reshape(Fdot, [], 1);
  A = reshape(Rgradient * fdot, 3, 3);
end
\end{verbatim}

%
%
%
%
%

\subsection{Metric}
\label{sec:pseudo_metric}

Let \verb!Sdot_1! through \verb!Sdot_m! be the $\dot\SS$ matrices of the $m$ tetrahedra of the $j^{th}$ representative node.
Then the liveliness metric is computed as:

\begin{verbatim}
metric_j = mean([ ...
  reshape(abs(Sdot_1), [], 1)
  ...
  reshape(abs(Sdot_m), [], 1)
	]);
\end{verbatim}

\subsection{Adjusted Jacobian}
\label{sec:pseudo_adjusted}

Let \verb!iq! and \verb!id! be the indices of the quasistatic and dynamic nodes, respectively, and \verb!nd=length(id)!.
Eqs.~\ref{eq:Jb} and \ref{eq:JA} are computed as:

\begin{verbatim}
% Jacobian without adaptivity
Kqq = K(iq,iq);
Kqd = K(iq,id);
J(iq,:) = -Kqq \ Kqd; % Eq. 6
J(id,:) = eye(nd); % Eq. 7

% Jacobian with adaptivity
KA = K;
KA(:,id) = 0; % zero out dynamic rows
KA(id,:) = 0; % zero out dynamic columns
KA = KA - sparse(id,id,ones(nd,1),n,n);
KqdA = K(:,id);
KqdA(id,:) = speye(nd);
JA = -KA \ KqdA; % Eq. 21
\end{verbatim}

\editEnd{}



\ifCLASSOPTIONcaptionsoff
  \newpage
\fi



\bibliographystyle{IEEEtran}
\bibliography{IEEEabrv,conjac}
\end{document}